\documentclass{jfm}

\usepackage{graphicx}
\usepackage{natbib}
\usepackage{amsmath}
\usepackage[version=3]{mhchem}
\usepackage{color}
\ifCUPmtlplainloaded \else
  \checkfont{eurm10}
  \iffontfound
    \IfFileExists{upmath.sty}
      {\typeout{^^JFound AMS Euler Roman fonts on the system,
                   using the 'upmath' package.^^J}%
       \usepackage{upmath}}
      {\typeout{^^JFound AMS Euler Roman fonts on the system, but you
                   dont seem to have the}%
       \typeout{'upmath' package installed. JFM.cls can take advantage
                 of these fonts,^^Jif you use 'upmath' package.^^J}%
       \providecommand\upi{\pi}%
      }
  \else
    \providecommand\upi{\pi}%
  \fi
\fi

\ifCUPmtlplainloaded \else
  \checkfont{msam10}
  \iffontfound
    \IfFileExists{amssymb.sty}
      {\typeout{^^JFound AMS Symbol fonts on the system, using the
                'amssymb' package.^^J}%
       \usepackage{amssymb}%
         \let\leq=\leqslant
         \let\geq=\geqslant
      }{}
  \fi
\fi

\ifCUPmtlplainloaded \else
  \IfFileExists{amsbsy.sty}
    {\typeout{^^JFound the 'amsbsy' package on the system, using it.^^J}%
     \usepackage{amsbsy}}
    {\providecommand\boldsymbol[1]{\mbox{\boldmath $##1$}}}
\fi

\providecommand\bnabla{\boldsymbol{\nabla}}
\providecommand\bcdot{\boldsymbol{\cdot}}
\newcommand\thb{\boldsymbol{\theta}}
\newcommand\phb{\boldsymbol{\phi}}
\newcommand\omb{\boldsymbol{\omega}}
\newcommand\rrb{\boldsymbol{r}}
\newcommand\xxb{\boldsymbol{x}}
\newcommand\yyb{\boldsymbol{y}}
\newcommand\zzb{\boldsymbol{z}}
\newcommand\nnb{\boldsymbol{n}}
\newcommand\kkb{\boldsymbol{k}}
\newcommand\vvb{\boldsymbol{v}}
\newcommand\VVb{\boldsymbol{V}}
\newcommand\bbb{\boldsymbol{b}}
\newcommand\IIb{\boldsymbol{I}}
\newcommand\Lin{\mathcal{L}}

\newcommand\vecth{\left( 
\begin{array}{c}
	\cos\theta \cos\phi\\
	\cos\theta \sin\phi\\
	-\sin\theta\\
\end{array}
\right)}
\newcommand\vecph{\left( 
\begin{array}{c}
	-\sin\phi\\
	\cos\phi\\
	0\\
\end{array}
\right)}

\newcommand\Dam{\mbox{\textit{Da}}}  

\newsavebox{\astrutbox}
\sbox{\astrutbox}{\rule[-5pt]{0pt}{20pt}}

\newcommand\p{\ensuremath{\partial}}

\newcommand\ttz{\ensuremath{\rightarrow 0}}
\newcommand\tti{\ensuremath{\rightarrow\infty}}

\title[Artificial Chemotaxis of Phoretic Swimmers]{Artificial chemotaxis of phoretic swimmers: Instantaneous and long-time behaviour}

\author[Maria T\u{a}tulea-Codrean and Eric Lauga]%
{Maria T\u{a}tulea-Codrean and Eric Lauga%
  \thanks{Email address for correspondence: e.lauga@damtp.cam.ac.uk}}

\affiliation{Department of Applied Mathematics and Theoretical Physics, University of Cambridge,
Wilberforce Road, Cambridge CB3 0WA, UK}

\pubyear{2010}
\volume{650}
\pagerange{119--126}
\date{\today}
\begin{document}

\maketitle

\begin{abstract}
Phoretic swimmers are a class of artificial active particles that has received significant attention in recent years. By making use of self-generated gradients (e.g.~in temperature, electric potential or some chemical product) phoretic swimmers are capable of self-propulsion without the complications of mobile body parts or a controlled external field. Focusing on diffusiophoresis, we quantify in this paper the mechanisms through which phoretic particles  may achieve chemotaxis, both at the individual and the non-interacting population level.  We first  derive a fully analytical law for the instantaneous propulsion and orientation of a phoretic swimmer with general axisymmetric surface properties, in the limit of zero P\'{e}clet number and small Damk\"{o}hler number. We then apply our results to the case of a Janus sphere, one of the most common designs of phoretic swimmers used in experimental studies.  We next put forward a novel application of generalised Taylor dispersion theory in order to characterise the long-time behaviour of a population of non-interacting phoretic swimmers. We compare our theoretical results with  numerical simulations for the mean drift and anisotropic diffusion of phoretic swimmers in chemical gradients. Our results will help inform the design of phoretic swimmers in future experimental applications.
\end{abstract}
 

\section{Introduction}

The academic community has recently taken significant interest in the better understanding of the locomotion of  microorganisms \citep{Lauga2009, Lauga2016} and the design of biomimetic devices \citep{Nelson2010}. One popular type of synthetic swimmers are  those able to self-propel through autophoresis -- a type of design that escapes the technical challenges associated with manufacturing motile body parts on a small scale. This class of devices makes use of self-generated gradients in temperature (thermophoresis), electric potential (electrophoresis) or concentration of some chemical species (diffusiophoresis) in order to induce motion. While currently at the centre of active research, the physical ideas behind autophoresis  are by no means novel and a comprehensive overview of the theory of phoretic transport can be found in the classical   article by \citet{Anderson89}. 

Recent technological advances have facilitated the manufacturing of diffusiophoretic swimmers in a variety of shapes and chemical properties, starting with pioneering experimental work done by \citet{paxton2004} and followed by a suite of detailed experiments  \citep[][]{Howse2007,Ebbens2011,ebbens2012,ebbens2014}. There has also been much interest recently in the study of thermophoretic particles \citep[][]{jiang2010,golestanian2012,bickel2013} as well as in the understanding of Marangoni self-propelled droplets \citep[][]{thutupalli2011,schmitt2013,izri2014}. 

In the present paper, we focus on the mechanism of diffusiophoresis due to concentrations of non-ionic species. Many recent papers investigate ways of exploiting geometric asymmetry \citep[][]{shklyaev2014, Lauga15}, asymmetry of surface properties \citep[][]{golestanian2005, Golestanian07} or a combination of the two \citep[][]{popescu2010} in order to generate a local imbalance of concentration in an otherwise uniform medium and induce motion. 

In this paper we  quantify the mechanisms through which  diffusiophoretic particles may achieve artificial chemotaxis, i.e.~the directed motion along an external chemical gradient, both at the individual and the non-interacting population level. It is  known that an asymmetric swimmer placed in a uniform background gradient will undergo active reorientation and experience a torque that seeks to align its axis of symmetry with the direction of the gradient \citep{bickel2014,Saha2014}.   Although the mechanisms through which diffusiophoretic swimmers achieve chemotaxis are qualitatively different from the techniques used by living organisms \citep[see][]{Berg75}, the prospect of achieving the same functionality is highly desirable for both biomedical and technological applications \citep{Nelson2010, popescu2011}.

In this work we first quantify the propulsion and reorientation mechanisms associated with the canonical problem of a spherical axisymmetric swimmer placed in a uniform background gradient of solute concentration. We approach this problem using the classical continuum framework of diffusiophoresis \citep{Golestanian07,julicher2009,sabass2012} as opposed to the osmotic framework proposed by Brady and coworkers \citep{cordova2008,brady2011,cordova2013}. To the best of our knowledge, we use the same setup as \citet{Saha2014}, but we generalise and correct their results. 
Specifically, we  derive a fully analytical law for the instantaneous propulsion and orientation of a phoretic swimmer with general axisymmetric surface properties, in the limit of zero P\'{e}clet number for both substrate and product. We compute the solution for a weakly-reactive swimmer as an expansion in small   Damk\"{o}hler number, including the first-order effects from the substrate which had been neglected by \citet{Saha2014}. The advantage of having a complete analytical law is that we are able to apply our results to real-life designs of phoretic swimmers, such as the Janus sphere, which could not be achieved using the truncated results previously published.

We next develop a continuum model for the long-time behaviour of a population of non-interacting phoretic swimmers. Most practical applications rely on the synchronised behaviour of a large number of phoretic swimmers \citep{Wang2013,palacci2013,palacci2014} and macroscopic models are essential in understanding how these micro-swimmers would disperse under the competing action of active propulsion and thermal fluctuations.  Our approach is inspired by work done on gyrotactic micro-organisms which successfully explained the formation of bioconvection patterns in cells such as \textit{Chalmydomonas nivalis}. Specifically, we use the framework of generalised Taylor dispersion theory developed by \citet{Frank91, Frank93} and apply it to the artificial chemotaxis of phoretic swimmers. We finally compare the predictions of our continuum model with numerical simulations of large samples of swimmers and obtain excellent agreement. Notably, our continuum model demonstrates  that the interplay between active propulsion and reorientation leads to the anisotropic diffusion of phoretic swimmers, a piece of physical understanding  overlooked by previous studies.
 
\section{Instantaneous behaviour: Artificial chemotaxis of an individual particle}
\label{section-2}
In this first part of the paper, we derive the instantaneous law controlling the linear and angular transport  of a phoretically-active sphere in a solute (or ``substrate'') gradient.

\subsection{Setup: Chemical problem}

A spherical particle of radius $R$ is immersed in a uniform gradient of substrate concentration and induces a disturbance to the equilibrium concentration of substrate not only due to the volume that it displaces, but also due to the chemical reaction that it facilitates at its surface. Before introducing the particle, the medium is characterised by a known linear background concentration of substrate,  $s_b(\rrb') = \rrb' \bcdot \bnabla s^{\infty}$, where $\rrb'$ is the position vector relative to some location of vanishing substrate concentration. The presence of the device leads to a perturbed concentration of substrate, $s$, which must be calculated.

Due to a catalytic coating, we assume that the swimmer promotes the conversion of substrate molecules $S$ into product molecules $P$, after an intermediary stage of binding to the enzyme molecules $E$. The classical model for this type of chemical behaviour is referred to as Michaelis-Menten kinetics, and starts from the reaction formula
\begin{equation}
\ce{{E+S} <=>[k_1][k_{-1}] ES ->[k_2] {E+P}},
\end{equation}
where $k_1, k_{-1}$ and $k_2$ are reaction rates \citep[for a review, see][]{nelson2008}.
 
When the supply of substrate is much larger than the supply of enzyme, the reaction is limited by the availability of enzymatic sites, and the system classically reaches a steady non-equilibrium state. The rate of conversion of substrate molecules S into product molecules P is given by the Michaelis-Menten rule
\begin{equation}
\kappa(s) = \frac{\kappa_1 \kappa_2 s}{\kappa_2 + \kappa_1 s},
\label{eq-MM}
\end{equation} 
where $s$ is the substrate concentration, $\kappa_1 = k_1k_2/(k_{-1}+k_2)$ and $\kappa_2=k_2$. Note that $\kappa(s)$ is the rate of reaction per enzyme molecule, so the total flux at the surface of the swimmer will be the reaction rate $\kappa(s)$ multiplied by a measure of surface activity $\sigma(\theta,\phi)$, which quantifies the abundance of enzyme molecules at a particular point on the surface of the particle, parameterised by the spherical coordinate angles $(\theta,\phi)$ in a body-fixed frame of reference.

\begin{figure}
	\begin{minipage}[b]{0.5\textwidth}
		\centering
		\includegraphics[width=\textwidth]{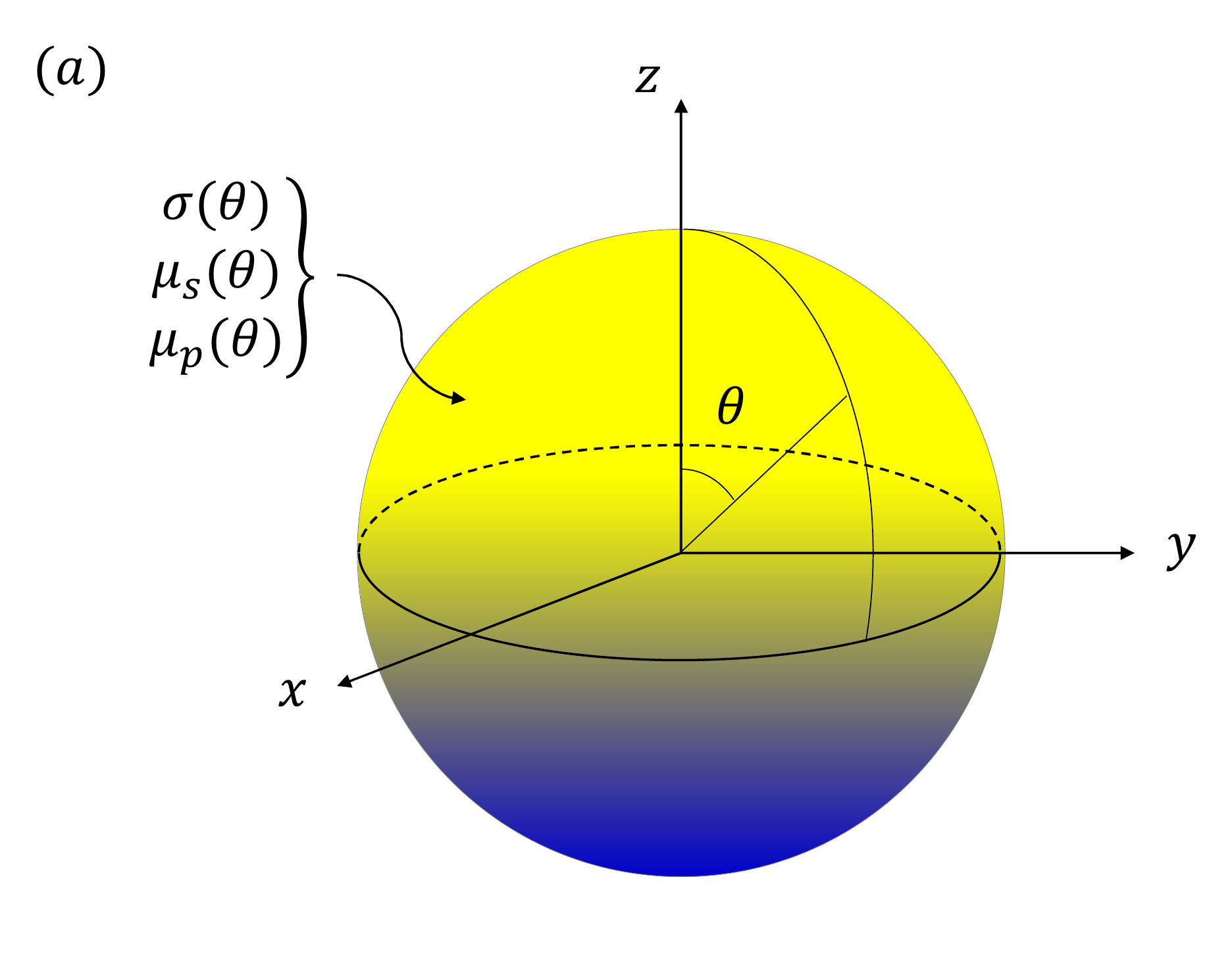}
	\end{minipage}
	\begin{minipage}[b]{0.5\textwidth}
		\centering
		\includegraphics[width=\textwidth]{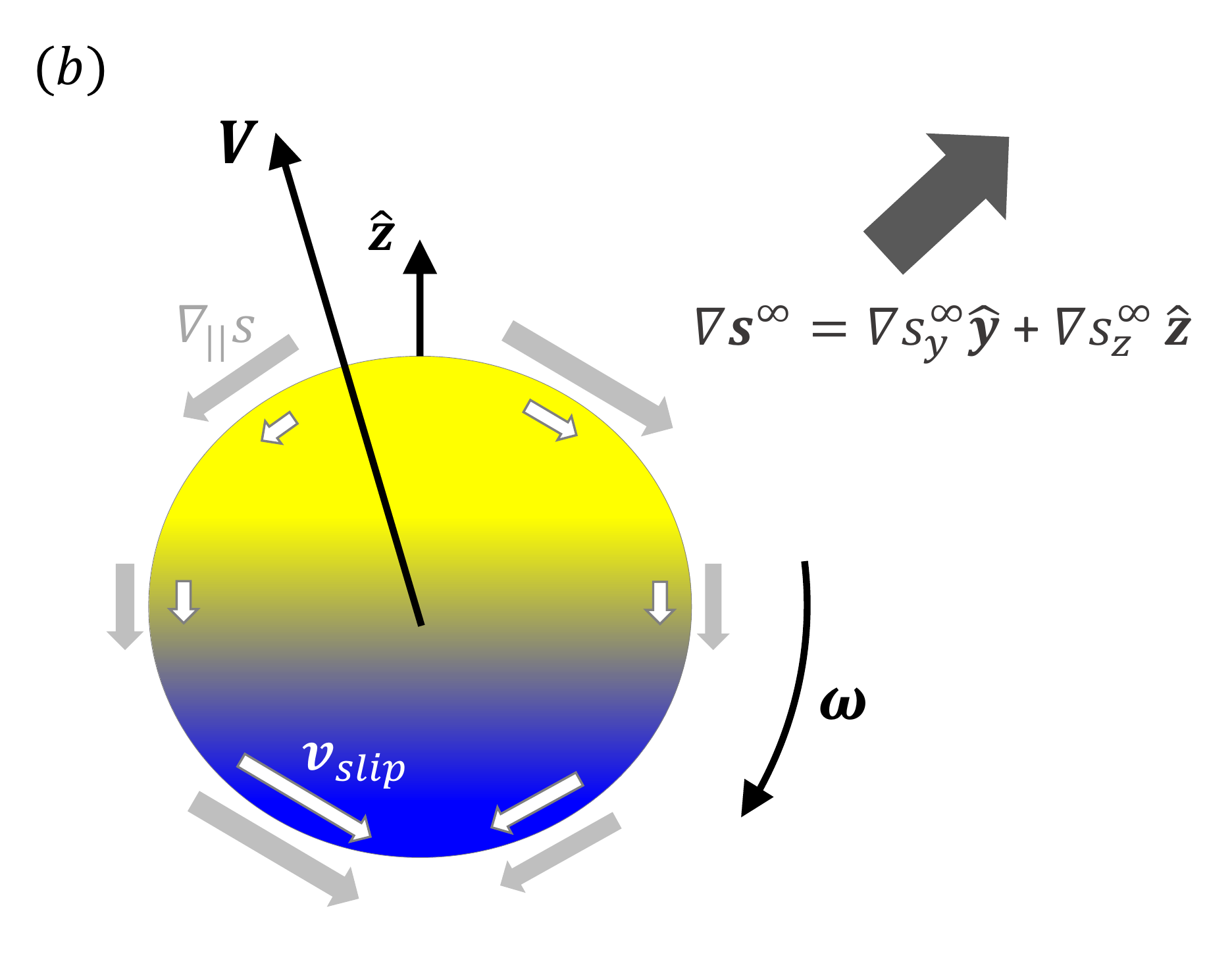}
	\end{minipage}
\caption{Sketch of the problem. 
(a): Axisymmetric active particle with axis of symmetry along       $z$. The surface properties (surface activity $\sigma$, substrate mobility $\mu_s$, and product mobility $\mu_p$) are functions of the polar angle $\theta$ alone. 
(b): Active particle  placed in a uniform substrate gradient, $\bnabla s^{\infty}$, which lies in the $yz$-plane. The presence of the background gradient, together with the catalytic coating of the swimmer, induce local tangential chemical gradients on the surface of the sphere (light grey arrows) which give rise to an apparent fluid slip velocity (white arrows) leading to translation (linear velocity $\VVb$) and rotation (angular velocity $\omb$) of the particle.}
\label{fig1}
\end{figure}

We next assume that the spherical swimmer has axisymmetric surface properties, and define the $z$-axis to be its axis of symmetry (see figure~\ref{fig1}a). We choose the origin at the centre of the sphere, and define our spherical coordinate system such that the polar angle $\theta$ is measured relative to the axis of symmetry of the swimmer. In this coordinate system, any  surface property of the swimmer can be expressed as series of Legendre polynomials of the polar angle $P_l(\cos\theta)$. In particular, the axisymmetric surface activity $\sigma(\theta)$ can be written as 
\begin{equation}
\sigma(\theta) = \sum_{l=0}^{\infty} \sigma_l P_l(\cos\theta).
\end{equation}

We note that the typical scale of substrate concentration surrounding a swimmer centred at position $\rrb_c$ is given by the background concentration of substrate at that position, i.e. $s \sim s_b(\rrb_c)$. It is this scale that determines the magnitude of the reaction rate on the surface of the swimmer. Using this scale, the nonlinear reaction rate from \eqref{eq-MM} has two distinguished limits in which analytical progress can be made. 

In the case when $s_b(\rrb_c) \gg \kappa_2/\kappa_1$, the reaction saturates at a constant rate $\kappa_2$. This is the limit studied by  \citet{golestanian2005,Golestanian07,popescu2010} and \citet{Lauga15}, amongst others. The swimmer could still perform chemotaxis in this regime but solely due to the substrate  since the product problem has become independent of the substrate dynamics. If the product is released at a constant rate $\kappa_2 \sigma(\theta)$ from the surface of the swimmer, it can only lead to translation along the axis of symmetry of the swimmer. Therefore, there is no chemotactic effect from the product in this regime.

In contrast,  when $s_b(\rrb_c) \ll \kappa_2/\kappa_1$, the reaction rate becomes proportional to the substrate concentration, 
\begin{equation}
\kappa(s) \sim \kappa_1 s,
\label{eq-linear-regime}
\end{equation}
which means that the release of product molecules at the surface of the swimmer   also depends on the direction of the background gradient in substrate concentration. In this limit, both substrate and product can contribute to chemotaxis in a non-trivial way. 

In what follows we shall focus on the linear regime \eqref{eq-linear-regime} which provides a mathematically tractable problem yet one that   leads to interesting chemotactic effects. Traditional diffusiophoresis (whereby a passive colloid is propelled due to a background gradient in the concentration of a chemical species) will also be captured in our model as the special case where the surface activity $\sigma$ is zero, and the presence of chemical reactions will add further complexity to the chemotactic behaviour of the swimmer.

In the limit of zero P\'{e}clet number for both substrate and product, we can neglect the advection of molecules and solve a diffusion-reaction problem with the appropriate boundary conditions. In the bulk of the fluid, we must solve the steady diffusion equation for each component
\begin{equation}
D_s \nabla^2 s = 0, \quad D_p \nabla^2 p = 0,
\label{eq-setup-diff}
\end{equation}
where $s$ and $p$ are the volume concentrations of substrate and product molecules, while $D_s$ and $D_p$ are their respective diffusivities.   The diffusive flux normal to the particle surface is given by the consumption and production of substrate and product molecules, respectively. Given the definition of the surface activity, $\sigma$,  we thus write these boundary conditions as 
\begin{equation}
\left. -D_s \frac{\p s}{\p n} \right|_{|\rrb| = R} = \left.-\kappa_1 \sigma(\theta) s\right|_{|\rrb| = R}, \quad \left. -D_p \frac{\p p}{\p n} \right|_{|\rrb| = R} = \left.\kappa_1 \sigma(\theta)s\right|_{|\rrb| = R}.
\label{eq-chem-BC}
\end{equation}
Finally, we impose that the perturbation due to the presence of the particle decays at infinity, such that we recover the background concentrations far away from the particle. Therefore, as $|\rrb| \rightarrow \infty$, we impose that
\begin{equation}
s \rightarrow s_b(\rrb_c) + \rrb \bcdot \bnabla s^{\infty}, \quad p \ttz, 
\label{eq-chem-inf}
\end{equation} 
where $\rrb_c$ is the instantaneous position of the centre of the swimmer relative to the reference location where $s_b (\rrb') = 0$,  and $\rrb = \rrb' - \rrb_c$ is the position relative to the centre of the swimmer.

\subsection{Setup: Hydrodynamics} 
\label{sec-hydro} 

The physical basis for the self-propulsion of phoretic swimmers lies in the interactions between the chemical molecules present in the fluid and the surface of the particle (see sketch in figure~\ref{fig1}b with notation). Following the classical framework \citep{Anderson89},  interactions are restricted to a thin diffuse layer around the particle where the chemical molecules obey a Boltzmann distribution. The energy of the particle is given by a total interaction potential $\Phi$ which includes effects such as van der Waals forces and excluded volume effects. We assume that the size of the boundary layer is much smaller than the inverse of the local curvature everywhere on the surface of the particle, so that we can use a locally flat approximation.

If there is a gradient in the concentration of the chemical species outside the diffuse boundary layer, it will give rise generically to an osmotic flow near the surface of the swimmer. This flow will lead to an apparent slip velocity at the edge of the diffuse boundary layer proportional to the tangential gradient in concentration of the chemical species. The constant of proportionality is called the diffusio-osmotic mobility, $\mu$, and it is a property of the surface interactions. A detailed description of diffusiophoresis can be found in \citet{Anderson89}, accompanied by the classical integral expression for the mobility $\mu$ in terms of the total interaction potential $\Phi$. For the purpose of the present paper, we may ignore the details of the interaction potential and simply work with the mobility as a given surface property.

Because our swimmer is axisymmetric, we can also express the diffusio-osmotic mobility as a series of Legendre polynomials, as we did for the surface activity. In general   there are two relevant mobilities, one for the substrate molecules ($\mu_s$) and the other for the product molecules ($\mu_p$), both of which we decompose as 
\begin{equation}
\mu_s(\theta) = \sum_{l=0}^{\infty} \mu_{sl} P_l(\cos\theta), \quad \mu_p(\theta) = \sum_{l=0}^{\infty} \mu_{pl} P_l(\cos\theta).
\end{equation}

In the case of our spherical particle, the slip velocity due to either one of the chemical species at a point $(\theta,\phi)$ on the surface of the swimmer can then be written as 
\begin{equation}
\vvb^\alpha_{slip} (\theta,\phi) = \mu_\alpha(\theta) (\IIb-\hat{\rrb}\hat{\rrb})\bcdot\bnabla \alpha |_{r=R}(\theta,\phi),
\label{eq-slip-velocity}
\end{equation}
where $\mu_\alpha$ is the diffusio-osmotic mobility of the surface with respect to the chemical species in question, $(\IIb-\hat{\rrb}\hat{\rrb})\bcdot\bnabla$ is the tangential gradient operator on the surface of the sphere, and $\alpha$ is the concentration of the chemical species of interest (substrate or product).

The final step in the setup of the problem is to translate the local slip velocity into the equations of motion of the particle. Within the low Reynolds number regime, inertia can be neglected and therefore it is relevant to talk about the instantaneous linear and angular velocities of the particle. A classical calculation using the reciprocal theorem on a force-free, torque-free sphere \citep[see][]{Stone96} establishes that the linear velocity is given by
\begin{equation}
\VVb^\alpha = - \frac{1}{4\pi} \int_0^{2\pi} \int_0^{\pi} \vvb^\alpha_{slip}(\theta,\phi)\sin\theta\mathrm{d}\theta \mathrm{d}\phi,
\label{eq-linear-velocity}
\end{equation}
and the angular velocity by
\begin{equation}
\omb^\alpha = - \frac{3}{8\pi R} \int_0^{2\pi} \int_0^{\pi} \hat{\rrb} \wedge \vvb^\alpha_{slip}(\theta,\phi)\sin\theta\mathrm{d}\theta \mathrm{d}\phi.
\label{eq-angular-velocity}
\end{equation}
Using the short-hand $\langle \dots \rangle$ to denote the surface average, we can write these as 
\begin{equation}
\VVb^\alpha = - \langle\vvb^\alpha_{slip}\rangle, \quad \omb^\alpha = -\frac{3}{2R} \langle \hat{\rrb} \wedge \vvb^\alpha_{slip}\rangle.
\label{eq-both-velocities}
\end{equation}

Due to the linearity of Stokes flow and of the chemical problem, the contributions from substrate (s) and product (p) can be calculated separately and added up at the end to give the total linear and angular velocities as
	\begin{equation}
	\VVb = \VVb^s + \VVb^p, \quad \omb = \omb^s + \omb^p.
	\end{equation}
In the simple case of a particle with uniform surface mobility and no surface activity, the only variations in concentration are due to the linear background gradient. If the interaction potential between the  surface of the particle and the chemical species were attractive (resp.~repulsive), then this would gives rise to a negative (resp.~positive) mobility \citep[see][]{Anderson89}. Thus, the osmotic flows induced near the surface would go against (or up) the tangential chemical gradient. According to   \eqref{eq-linear-velocity} this means that the particle in this case would move in the direction of increasing (resp.~decreasing) chemical concentration, as  expected from a colloid which displays attractive (resp.~repulsive) interactions with the chemical species. This is the physical basis for the chemotaxis of phoretic swimmers, and the following subsections are concerned with formulating a precise mathematical description of the chemotactic mechanism. 

\subsection{Small Damk\"{o}hler number approximation}

In its current form, the boundary condition \eqref{eq-chem-BC} at the surface of the sphere does not admit a closed-form solution for the substrate. In order to make further progress, we must make a final simplifying assumption about the relative importance of reaction and diffusion for the substrate molecules. This balance is quantified by the Damk\"{o}hler number, $\Dam$,  which we define as
	\begin{equation}
	\Dam = \frac{\kappa_1 \Sigma R}{D_s}  \sim \frac{|\kappa_1 \sigma(\theta) s|}{|D_s \p s/\p n|},
\end{equation}   
where $\kappa_1$ is the volumetric rate or reaction (in units $\mbox{L}^3\mbox{T}^{-1}$) and $\Sigma$ is the typical scale of surface activity (in units $\mbox{L}^{-2}$). In the limit of small Damk\"{o}hler number, $\Dam \ll 1$, the particle behaves as a passive sphere at leading order ($\Dam^0$) but, at first order ($\Dam^1$), it produces small disturbances in the substrate and product distributions due to the chemical reactivity of its surface.

To make this statement more rigorous, we proceed with non-dimensionalising the equations of the chemical problem. For the chemical concentrations we use the scale imposed by the given background concentration of substrate, such that our dimensionless variables are $\tilde{s}=s/s_b(\rrb_c)$ and $\tilde{p}=p/s_b(\rrb_c)$. We scale lengths by the radius of the swimmer, $\tilde{r}=r/R$, and the surface activity by its typical scale, $\tilde{\sigma} = \sigma/\Sigma$. Under these scalings, the dimensionless versions of equations \eqref{eq-setup-diff}-\eqref{eq-chem-inf} are
\begin{equation}
\nabla^2 \tilde{s} = 0, \quad \nabla^2 \tilde{p} = 0,
\end{equation}
in the bulk of the fluid, with boundary conditions on the surface of the swimmer
\begin{equation}
\left.\frac{\p \tilde{s}}{\p n}\right|_{\tilde{r}=1} 
= 
\Dam \tilde{\sigma}(\theta) \tilde{s}|_{\tilde{r}=1}, 
\quad
\left.\frac{\p \tilde{p}}{\p n}\right|_{\tilde{r}=1} 
= 
-\delta\Dam\tilde{\sigma}(\theta) \tilde{s}|_{\tilde{r}=1},
\label{eq-nondim-surf}
\end{equation}
 and  away from the swimmer
\begin{equation}
\tilde{s} \rightarrow 1 + \tilde{\rrb} \bcdot \bnabla \tilde{s}^{\infty}, \quad \tilde{p} \ttz ~~ \text{as} ~ \tilde{r}\tti.
\label{eq-nondim-inf}
\end{equation}
In addition to the Damk\"{o}hler number,   a second dimensionless parameter appears, $\delta = D_s/D_p$, which is the ratio of the two diffusivities.

Due to existence of a background substrate profile, the nondimensional substrate distribution around the swimmer will include an $\mathcal{O}(1)$ term, such that $\tilde{s}=\tilde{s}_0 + \Dam \tilde{s}_1 + \mathcal{O}(\Dam^2)$. This can be seen from the boundary conditions \eqref{eq-nondim-surf} and \eqref{eq-nondim-inf}. However, since there was no product in the ambient medium to begin with, the scale of the product distribution is determined by the chemical reactions at the surface of the swimmer. In the small Damk\"{o}hler number approximation the swimmer is only weakly reactive, which means that the leading-order product distribution appears only at $\mathcal{O}(\Dam)$, such that $\tilde{p}=\Dam \tilde{p}_1 + \mathcal{O}(\Dam^2)$.

Due to linearity of the Stokes equations, the contributions from substrate and product add up to give the total linear and angular velocities of the swimmer. Therefore, if we want to include all terms of equal importance for the motion of the swimmer, we must calculate the substrate and product concentrations to the same level of accuracy in our Damk\"{o}hler number expansion. In the current paper we work out the solution up to $\mathcal{O}(\Dam)$, which brings the first non-trivial contributions to the final answer and also captures the expected phenomenology.

Returning to dimensional variables, we will approximate the substrate concentration to be $s \simeq s_0 + s_1$, where $s_0$ is the solution of the leading-order problem
\begin{equation}
\nabla^2 s_0 = 0, \quad \left.\frac{\p s_0}{\p n}\right|_{r=R} = 0, \quad s_0 \rightarrow s_b(\rrb_c) + \rrb \bcdot \bnabla s^{\infty} ~ \text{as} ~ r\tti,
\label{eq-substrate-leading}
\end{equation}
while $s_1$ is the solution of the first-order problem 
\begin{equation}
\nabla^2 s_1 = 0, \quad \left.\frac{\p s_1}{\p n}\right|_{r=R} = \frac{\kappa_1  \sigma(\theta)}{D_s}\left.s_0\right|_{r=R}, \quad s_1 \ttz ~ \text{as} ~ r\tti.
\label{eq-substrate-first}
\end{equation}

Similarly, we approximate the product concentration to be $p \simeq p_1$, where $p_1$ is the solution of the first-order problem
\begin{equation}
\nabla^2 p_1 = 0, \quad \left.\frac{\p p_1}{\p n}\right|_{r=R} = -\frac{\kappa_1  \sigma(\theta)}{D_p}\left.s_0\right|_{r=R}, \quad p_1 \ttz ~ \text{as} ~ r\tti.
\label{eq-product-first}
\end{equation}

Notice that the first-order problems for the substrate and product are the same, up to a factor of $-D_s/D_p$, so we only need to solve for one of them.

\newpage
\subsection{Leading-order effects}
\label{subsec-leading}

Recall that our $z$-axis is the axis of symmetry of the swimmer. Without loss of generality, we may then choose our $y$-axis such that the substrate gradient lies in the $yz$-plane, that is $\bnabla s^\infty = \nabla s_y^\infty \hat{\yyb} + \nabla s_z^\infty \hat{\zzb}$. 

The leading-order substrate concentration can be written as the sum of the background concentration plus a perturbation which satisfies Laplace's equation and decays at infinity. The general solution to Laplace's equation can be written classically as a sum of spherical harmonics $Y_l^m(\theta,\phi) \propto e^{im\phi}P_l^m(\cos\theta)$, where $P_l^m$ is an associated Legendre polynomial, multiplied by an appropriate radial component. For our perturbation to decay at infinity this radial component has to be a multiple of $r^{-(l+1)}$, so we can write
\begin{equation}
s_0(r,\theta,\phi) = s_b (\rrb_c) + \mathbf{r} \bcdot \bnabla s^\infty + \sum_{l=0}^{\infty} \sum_{m=-l}^{l} f_l^m r^{-(l+1)}Y_l^m(\theta,\phi).
\end{equation}
This expression can be simplified further if we take into account that the zero-flux boundary condition from \eqref{eq-substrate-leading} can only excite the spherical harmonics of order $m=0$ and $m=\pm 1$, which are already present in the solution due to the form of the concentration at infinity. Due to our choice of $y$-axis, the $m=\pm 1$ components must only lead to dependence on $\sin\phi$ and not $\cos\phi$, so the substrate concentration has at most the form
\begin{equation}
s_0(r,\theta,\phi) = s_b (\rrb_c) + \mathbf{r} \bcdot \bnabla s^\infty + \sum_{l=0}^{\infty} \left(\frac{R}{r}\right)^{l+1}\left[a_l P_l(\cos\theta) + b_l P_l^1(\cos\theta)\sin\phi\right].
\end{equation}

Imposing the zero-flux boundary condition from \eqref{eq-substrate-leading}, we  find the coefficients
\begin{equation}
a_1 = \frac{\nabla s_z^\infty R}{2}, \quad b_1 = -\frac{\nabla s_y^\infty R}{2},
\end{equation}
and $a_l, b_l = 0$ for $l\neq 1$,
which lead to the following substrate distribution
\begin{equation}
s_0 = s_b (\rrb_c) + \rrb \bcdot \bnabla s^\infty + \frac{R^3}{2r^2}\left(\nabla s_z^\infty \cos\theta + \nabla s_y^\infty\sin\theta\sin\phi\right). 
\end{equation}
Notably, this expression  can be simplified into
\begin{equation}
s_0 = s_b (\rrb_c) + \rrb \bcdot \bnabla s^\infty \left( 1 + \frac{1}{2} \left(\frac{R}{r}\right)^3\right).
\label{eq-substrate-concentration} 
\end{equation}

In order to compute the slip velocity, $\vvb^{s,0}_{slip}$,  due to the leading-order substrate profile,  we have to evaluate 
\begin{equation}
(\IIb-\hat{\rrb}\hat{\rrb}) \bcdot \bnabla s_0 = \frac{1}{r} \frac{\p s_0}{\p \theta} \skew3\hat{\thb} + \frac{1}{r \sin\theta} \frac{\p s_0}{\p \phi} \skew3\hat{\phb}
\end{equation}
on the surface of the swimmer. Using \eqref{eq-slip-velocity}, we obtain
\begin{equation}
\vvb^{s,0}_{slip} = \frac{3\mu_s(\theta)}{2}  \left[(- \nabla s_z^\infty \sin\theta + \nabla s_y^\infty \cos\theta\sin\phi) \skew3\hat{\thb} + \nabla s_y^\infty \cos\phi \ \skew3\hat{\phb} \right].
\label{eq1-solute}
\end{equation}

Then, by averaging this slip velocity over the surface of the spherical swimmer, we determine the contribution to linear velocity, $\VVb^{s}_0$,  coming from the leading-order substrate problem  as
\begin{equation}
\VVb^{s}_0 = -\left(\mu_{s0}+\frac{1}{10}\mu_{s2}\right) \bnabla s^\infty + \frac{3}{10}\mu_{s2} \ \skew3\hat{\zzb}\skew3\hat{\zzb} \bcdot \bnabla s^\infty.
\label{eq-passive-trans}
\end{equation}
The calculations leading to this result are detailed in the first subsection of appendix \ref{appA}.

Although this result is valid for a general surface mobility, $\mu_s(\theta)$,   only two Legendre modes of the mobility  contribute to the final results, namely  those adjacent to the forcing coming from the linear gradient. Furthermore, our result is in   agreement with the linear velocity calculated by Anderson for a passive sphere placed in a non-uniform field \citep[equation (37a) from][]{Anderson89}. In our case that field is the concentration of substrate molecules, and the sphere is   passive, at leading order, due to our small Damk\"{o}hler number approximation.

Similarly, one has to compute the average of $\hat{\rrb} \wedge \vvb^{s,0}_{slip}$ over the surface of the spherical swimmer in order to determine the angular velocity contribution, $\omb^{s}_0$, resulting from the leading-order substrate problem
\begin{equation}
\omb^{s}_0 =  -\frac{3\mu_{s1}}{4R} \ \skew3\hat{\zzb} \wedge \bnabla s^\infty.
\label{eq-passive-ang}
\end{equation}
The calculations leading to this are detailed in the second subsection of appendix \ref{appA}.

This result agrees with the angular velocity calculated by Anderson for a passive sphere placed in a non-uniform field \citep[equation (37b) from][]{Anderson89}.
Note again that this result is valid for a general surface mobility $\mu_s(\theta)$, but only one mode contributes to the reorientation of a passive swimmer. This mode corresponds to a linear dependence of mobility on the cosine of the polar angle, with opposite and equal in magnitude mobilities at the two poles. By symmetry we would expect this type of swimmer to align with the chemical gradient, but we can also predict the direction of rotation using a simple physical argument.  Following the negative and positive chemotactic response of swimmers with uniform positive and negative mobility, respectively, as discussed at the end of section \S\ref{sec-hydro}, we would expect the pole of negative mobility to pull towards the higher concentration of substrate and the pole of positive mobility to pull towards the lower concentration. This  will then make the swimmer rotate until its axis of symmetry is aligned with the substrate gradient and the pole of negative mobility faces higher concentration,  which is  precisely the dynamics captured by   \eqref{eq-passive-ang}. Since the two poles are equally strong, the tug-of-war cannot be won by either opponent and this mode has zero linear velocity, in agreement with  \eqref{eq-passive-trans}.

\subsection{First-order effects}
\label{subsec-first}
We now turn our attention to finding the corrections to first order in Damk\"{o}hler number for the substrate problem. In order to satisfy Laplace's equation and the decay at infinity, the first-order substrate concentration must have the general form
\begin{equation}
s_1(r,\theta,\phi) = \sum_{l=0}^{\infty} \sum_{m=-l}^{l} F_l^m r^{-(l+1)}Y_l^m(\theta,\phi),
\end{equation}
but the normal-flux boundary condition from \eqref{eq-substrate-first} can only excite modes with either no $\phi$-dependence or $\sin\phi$-dependence which appear in the leading-order substrate concentration from   \eqref{eq-substrate-concentration}. This means that the first-order substrate concentration can be simplified at most to the sum
\begin{equation}
s_1(r,\theta,\phi) = \sum_{l=0}^{\infty} \left(\frac{R}{r}\right)^{l+1} \left[A_l P_l(\cos\theta) - B_l \sin\theta P'_l(\cos\theta) \sin\phi \right],
\label{eq-s-1-general-form}
\end{equation}
where we have used the fact that $P_l^1(\cos\theta) = -\sin\theta P'_l(\cos\theta)$. The coefficients $A_l$ and $B_l$ can be determined from the normal-flux boundary condition in \eqref{eq-substrate-first}. These calculations are detailed in the third subsection of appendix \ref{appA}, and lead to the following expressions
\begin{equation}
A_l = -\frac{\kappa_1 R}{(l+1)D_s} \left[s_b (\rrb_c) \sigma_l + \frac{3}{2} \nabla s_z^\infty R \left(\frac{l+1}{2l+3} \sigma_{l+1} + \frac{l}{2l-1}\sigma_{l-1}\right)\right],
\end{equation}
\begin{equation}
B_l = -\frac{3\kappa_1 R^2 \nabla s_y^\infty}{2(l+1)D_s} \left(\frac{\sigma_{l+1}}{2l+3} - \frac{\sigma_{l-1}}{2l-1} \right).
\end{equation}

Since the first-order product problem \eqref{eq-product-first} is the same as the first-order substrate problem \eqref{eq-substrate-first}, except for a factor of $-D_s/D_p$, we can use our results for $s_1$ and state that the first-order product concentration $p_1$ is given by
\begin{equation}
p_1 = \sum_{l=0}^{\infty} \left(\frac{R}{r}\right)^{l+1} \left[C_l P_l(\cos\theta) - D_l \sin\theta P'_l(\cos\theta) \sin\phi \right],
\end{equation}
where the coefficients $C_l$ and $D_l$ can be written as
\begin{eqnarray}
C_l &=& \frac{\kappa_1 R}{(l+1)D_p} \left[s_b (\rrb_c) \sigma_l + \frac{3}{2} \nabla s_z^\infty R \left(\frac{l+1}{2l+3} \sigma_{l+1} + \frac{l}{2l-1}\sigma_{l-1}\right)\right],\\
D_l &=& \frac{3\kappa_1 R^2 \nabla s_y^\infty}{2(l+1)D_p} \left(\frac{\sigma_{l+1}}{2l+3} - \frac{\sigma_{l-1}}{2l-1} \right).
\end{eqnarray}

We next need to  calculate the contribution of the first-order substrate concentration to the slip velocity. Using \eqref{eq-slip-velocity} and \eqref{eq-s-1-general-form}, we find that
\begin{multline}
\vvb^{s,1}_{\text{slip}} = \frac{\mu_s(\theta)}{R} \sum_{l=0}^{\infty} \left(- A_l\sin\theta P'_l(\cos\theta) -B_l \cos\theta P'_l(\cos\theta) \sin\phi\right. \\ + \left.B_l \sin^2\theta P''_l(\cos\theta)\sin\phi\right)\skew3\hat{\thb} + \frac{\mu_s(\theta)}{R} \sum_{l=0}^{\infty} \left(-B_l P'_l(\cos\theta)\cos\phi\right) \skew3\hat{\phb}. 
\end{multline}

The last two subsections in appendix \ref{appA} present the detailed calculations involved in averaging $\vvb^{s,1}_{\text{slip}}$ and $\hat{\rrb} \wedge \vvb^{s,1}_{\text{slip}}$ over the surface of the spherical swimmer, in order to calculate the contributions of the first-order substrate problem to the linear and angular velocities of the swimmer. The final results are that the first-order substrate problem contributes
\begin{multline}
\VVb^{s}_1 = -\frac{\kappa_1 s_b (\rrb_c)}{D_s} \sum_{l=1}^{\infty} \left(\frac{l}{2l+1}\right)\sigma_l \left(\frac{\mu_{s,l+1}}{2l+3}-\frac{\mu_{s,l-1}}{2l-1}\right) \skew3\hat{\zzb} \\
- \frac{3 \kappa_1 R}{4 D_s}\sum_{l=1}^{\infty} \left(\frac{l}{2l+1}\right)\left(\frac{\sigma_{l+1}}{2l+3} - \frac{\sigma_{l-1}}{2l-1} \right)\left(\frac{l+1}{2l-1}\mu_{s,l-1}+\frac{l}{2l+3}\mu_{s,l+1}\right) \bnabla s^\infty \\
+ \frac{3 \kappa_1 R}{4 D_s}\sum_{l=1}^{\infty} \left(\frac{l}{2l+1}\right)\left(\frac{3(l+1)\sigma_{l+1}\mu_{s,l-1}}{(2l+3)(2l-1)}-\frac{(l+2)\sigma_{l+1}\mu_{s,l+1}}{(2l+3)^2}\right. \\ +\left. \frac{(l-1)\sigma_{l-1}\mu_{s,l-1}}{(2l-1)^2} - \frac{3l\sigma_{l-1}\mu_{s,l+1}}{(2l-1)(2l+3)}\right) \skew3\hat{\zzb}\skew3\hat{\zzb}\bcdot\bnabla s^\infty.
\end{multline}
to the linear velocity of the swimmer, and
\begin{equation}
\omb^{s}_1= -\frac{9\kappa_1}{8 D_s} \sum_{l=1}^{\infty} \left(\frac{l}{2l+1}\right)\mu_{sl} \left(\frac{\sigma_{l+1}}{2l+3}-\frac{\sigma_{l-1}}{2l-1}\right) ~ \skew3\hat{\zzb} \wedge \bnabla s^\infty 
\end{equation}
to its angular velocity.

The contributions from the first-order product problem will be the same as the ones above, with an additional factor of $-D_s/D_p$ which carries through from the solution of the chemical problem, and a change of subscript on the mobility coefficients $\mu_{sl} \mapsto \mu_{pl}$.

\subsection{Summary of results}
In summary, we may write the instantaneous linear and angular velocities of the phoretic swimmer, when its centre is at position $\rrb_c$, in the compact form  
\begin{eqnarray}
\VVb &=& U(\rrb_c) \skew3\hat{\zzb}+\alpha \bnabla s^\infty + \beta \ \skew3\hat{\zzb}\skew3\hat{\zzb}\bcdot\bnabla s^\infty,\label{summary-linear}\\
\omb &=& \Phi \ \skew3\hat{\zzb} \wedge \bnabla s^\infty \label{summary-angular} . 
\end{eqnarray} 
Due to the linearity of the Stokes equations, the total linear and angular velocities of the phoretic swimmer are simply the sum of contributions from the substrate and the product, at leading and first order. Using the results from the previous subsections, we deduce that the coefficients appearing in \eqref{summary-linear} and \eqref{summary-angular} are given by
\begin{equation}
U(\rrb_c) = \kappa_1 s_b (\rrb_c) \sum_{l=1}^{\infty} \left(\frac{l}{2l+1}\right)\sigma_l \left[\frac{1}{2l+3}\left(\frac{\mu_{p,l+1}}{D_p}-\frac{\mu_{s,l+1}}{D_s}\right)-\frac{1}{2l-1}\left(\frac{\mu_{p,l-1}}{D_p}-\frac{\mu_{s,l-1}}{D_s}\right)\right], 
\label{eqn-U}
\end{equation}
\begin{multline}
\alpha = -\left(\mu_{s0}+\frac{1}{10}\mu_{s2}\right) + \frac{3\kappa_1 R}{4}\sum_{l=1}^{\infty} \left(\frac{l}{2l+1}\right)\left(\frac{\sigma_{l+1}}{2l+3} - \frac{\sigma_{l-1}}{2l-1} \right) \\ \times\left[\frac{l+1}{2l-1}\left(\frac{\mu_{p,l-1}}{D_p}-\frac{\mu_{s,l-1}}{D_s}\right)\right. +\left.\frac{l}{2l+3}\left(\frac{\mu_{p,l+1}}{D_p}-\frac{\mu_{s,l+1}}{D_s}\right)\right], \label{eqn-alpha-nu}
\end{multline}
\begin{multline}
\beta = \frac{3\mu_{s2}}{10} - \frac{3\kappa_1 R}{4}\sum_{l=1}^{\infty} \left(\frac{l}{2l+1}\right) \left[\left(\frac{3(l+1)\sigma_{l+1}}{(2l+3)(2l-1)} + \frac{(l-1)\sigma_{l-1}}{(2l-1)^2}\right) \left(\frac{\mu_{p,l-1}}{D_p}-\frac{\mu_{s,l-1}}{D_s}\right) \right. \\ \left.  -\left(\frac{(l+2)\sigma_{l+1}}{(2l+3)^2} + \frac{3l\sigma_{l-1}}{(2l-1)(2l+3)}\right)\left(\frac{\mu_{p,l+1}}{D_p}-\frac{\mu_{s,l+1}}{D_s}\right)\right], \label{eqn-beta-mu}
\end{multline}
\begin{equation}
\Phi = -\frac{3\mu_{s1}}{4R} + \frac{9\kappa_1}{8} \sum_{l=1}^{\infty} \left(\frac{l}{2l+1}\right)\left(\frac{\mu_{pl}}{D_p}-\frac{\mu_{sl}}{D_s}\right) \left(\frac{\sigma_{l+1}}{2l+3}-\frac{\sigma_{l-1}}{2l-1}\right).\label{eqn-Omega}
\end{equation}

To the best of our knowledge, the setup and modelling assumptions used in deriving these results follow those used by \citet{Saha2014}. Our final equations, (\ref{eqn-U})-(\ref{eqn-Omega}), represent an extension of the instantaneous laws of motion published in \citet{Saha2014}, to a phoretic swimmer with the most general axisymmetric surface activity $\sigma(\theta)$ and surface mobilities $\mu_s(\theta)$, $\mu_p(\theta)$. It is also important to note that our solution includes the first-order contributions from the substrate, an effect which had been neglected by \citet{Saha2014}.

\subsection{Janus spherical swimmer}

\begin{figure}
    \centering
    \includegraphics[width=.5\textwidth]{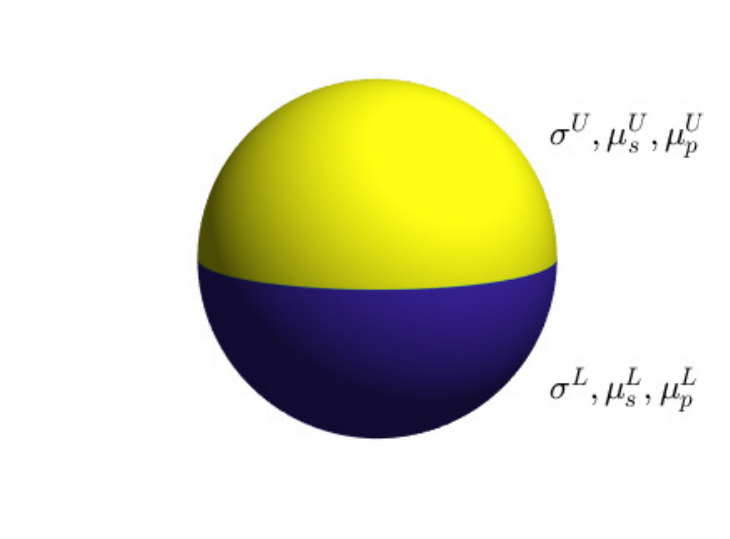}
    \caption{Schematic representation of a Janus spherical swimmer with surface properties $\{\sigma^U,\mu_s^U,\mu_p^U\}$ on the upper hemisphere and $\{\sigma^L,\mu_s^L,\mu_p^L \}$ on the lower hemisphere. The linear and angular velocities of this phoretic swimmer are given by    equations \eqref{eq-Janus-1}-\eqref{eq-Janus-4}.}
    \label{fig:Janus}
\end{figure}

Because our extended laws of motion allow for arbitrary axisymmetric surface properties, we can  apply our results to one of the most common designs of phoretic swimmers, namely that of a  Janus sphere. This type of swimmer, depicted in figure~\ref{fig:Janus}, has uniform surface properties on each of its two hemispheres such that the surface activity is given by
\begin{equation}\label{Jsigma}
\sigma(\theta) = \left\{
    \begin{array}{ll}
       \sigma^U, & 0 \leq \theta < \pi/2, \\[2pt]
       \sigma^L, & \pi/2 < \theta \leq \pi,
    \end{array} \right.
\end{equation}
where the superscript denotes the lower ($L$) and upper ($U$) halves of the swimmer. We also assume similar spatial distributions of the two mobilities $\mu_s(\theta)$ and $\mu_p(\theta)$ (see full notation in figure~\ref{fig:Janus}). 

With the help of the integral identity \eqref{eq-half-integral}, we can project \eqref{Jsigma} onto the space of Legendre polynomials, and we find that
\begin{equation}
\sigma_0 = \frac{1}{2}(\sigma^U+\sigma^L),
\end{equation}
but that all other even modes vanish
\begin{equation}
\sigma_{2k} = 0, \quad k \geq 1.
\end{equation}
For the odd modes, we have
\begin{equation}
\sigma_{2k+1}=(-1)^{k}(\sigma^U-\sigma^L)\frac{(4k+3)(2k+1)!!}{(4k+2)(2k+2)!!},
\end{equation}
and equivalent identities for the mobility coefficients $\mu_{sl}$ and $\mu_{pl}$.

Upon substituting these expressions into our results for the linear and angular velocity of a general axisymmetric swimmer,  we find that the parameters characterising the motion of a Janus sphere in a uniform chemical gradient,  (\ref{eqn-U})-(\ref{eqn-Omega}), are given by
\begin{eqnarray}
U_{\text{Janus}}&=&\frac{\kappa_1 s_b}{8}(\sigma^L-\sigma^U)\left(\frac{\mu_p^L+\mu_p^U}{D_p}-\frac{\mu_s^L+\mu_s^U}{D_s} \right),
\label{eq-Janus-1}\\
\alpha_{\text{Janus}} &=& -\frac{1}{2}(\mu_s^U+\mu_s^L)-\frac{\kappa_1 R}{8}(\sigma^U+\sigma^L)\left(\frac{\mu_p^L+\mu_p^U}{D_p}-\frac{\mu_s^L+\mu_s^U}{D_s} \right)\nonumber \\
&~& -\frac{3\gamma\kappa_1 R}{4}(\sigma^U-\sigma^L)\left(\frac{\mu_p^U-\mu_p^L}{D_p}-\frac{\mu_s^U-\mu_s^L}{D_s} \right), \quad\quad\\
\beta_{\text{Janus}} &=& -\frac{3\gamma\kappa_1 R}{4}(\sigma^U-\sigma^L)\left(\frac{\mu_p^U-\mu_p^L}{D_p}-\frac{\mu_s^U-\mu_s^L}{D_s} \right),\\
\Phi_{\text{Janus}}&=&\frac{9}{16R}(\mu_s^L-\mu_s^U)+\frac{9\kappa_1}{64}(\sigma^L+\sigma^U)\left(\frac{\mu_p^L-\mu_p^U}{D_p}-\frac{\mu_s^L-\mu_s^U}{D_s} \right),
\label{eq-Janus-4}
\end{eqnarray}
with swimming kinematics as in \eqref{summary-linear} and \eqref{summary-angular}. 
Note that the coefficients $\alpha_{\text{Janus}}$ and $\beta_{\text{Janus}}$ are more easily calculated from equation  \eqref{eq-needed-for-janus} than from equations \eqref{eqn-alpha-nu}-\eqref{eqn-beta-mu}, and that they involve a numerical factor
\begin{equation}
\gamma = \sum_{k=1}^\infty k\left(\frac{(2k-1)!!}{(2k-1)(2k+2)!!}\right)^2 \approx 0.0336.
\end{equation}
Our result for $U_{\text{Janus}}$ agrees with \citet{Golestanian07} if we equate their concentration-independent surface activity $\alpha(\theta)$ to the component of our surface activity which is independent of the background gradient, i.e.~$\kappa_1 s_b \sigma(\theta)$. Instead of a single chemical mobility to diffusivity ratio $\mu/D$, we also have an effective value, $(\mu/D)_{\text{eff}} = \mu_p/D_p - \mu_s/D_s$, due to the interplay of the two chemical species in our problem.
 
The simple expressions in \eqref{eq-Janus-1}-\eqref{eq-Janus-4} provide an estimate for the linear and angular velocity of the swimmer based on information about its surface properties alone, and could thus be used in the design and fabrication of phoretic Janus particles.  
 
Note that, at leading order, the only non-zero parameters are $\alpha_{\text{Janus}}$ and $\Phi_{\text{Janus}}$. Thus, the dominant behaviour of a spherical Janus swimmer consists in a constant linear velocity along the direction of the chemical gradient, with the orientation of the swimmer being irrelevant. In order to fabricate an efficient chemotactic swimmer of this type, one must simply ensure that $\mu_s^U+\mu_s^L$ has the desired sign for positive or negative chemotaxis. The simplest realisation of this is a swimmer with uniform mobility. However, if it happens that $\mu_s^U+\mu_s^L=0$ then the chemotactic behaviour of the Janus sphere will depend on first-order effects coming from the chemical reactions. At the end of this subsection we discuss the different possible strategies for chemotaxis.

For a swimmer that responds to both substrate and product molecules it is possible, in theory, to choose the surface properties $\{\sigma^L,\sigma^U,\mu_s^L,\mu_s^U,\mu_p^L,\mu_p^U \}$ in order to obtain any set of prescribed values for $\{ U_{\text{Janus}},\alpha_{\text{Janus}},\beta_{\text{Janus}},\Phi_{\text{Janus}}\}$, although this may be hard to achieve in practice. Surprisingly, the same cannot be said for a swimmer that responds only to product molecules and has $\mu_s =0$ even though there are four degrees of freedom and four target parameters. This is due to the specific combinations in which the four degrees of freedom appear in the expressions for $\{ U_{\text{Janus}},\alpha_{\text{Janus}},\beta_{\text{Janus}},\Phi_{\text{Janus}}\}$. We find that a Janus sphere with $\mu_s =0$ is subject to the following dependence relation between the four parameters describing its motion
\begin{equation}
    \alpha_{\text{Janus}}=\frac{16\gamma R^2 U_{\text{Janus}}\Phi_{\text{Janus}}}{3s_b \beta_{\text{Janus}}}+\beta_{\text{Janus}}.
\end{equation}
This represents a constraint on the range of behaviour that can be obtained with a swimmer that has zero mobility with respect to the substrate molecules.

For a swimmer that responds to the product molecules, and possibly the substrate molecules as well, we can distinguish different strategies for chemotaxis depending on the quantitative details of the problem. If the chemical gradient is sufficiently strong, the linear velocity of the swimmer will be dominated by the chemotactic `sedimentation' terms, $\alpha \bnabla s^\infty + \beta \ \skew3\hat{\zzb}\skew3\hat{\zzb}\bcdot\bnabla s^\infty$. In this case it is better to opt for a simpler swimmer with uniform surface properties, which ensures that $\beta_{\text{Janus}}=0$ and the swimmer undergoes direct chemotactic sedimentation along the gradient, with the sign of $\alpha_{\text{Janus}}$  determining the sense of chemotaxis. 

If, on the other hand, the chemical gradient is sufficiently weak, then the linear velocity of the swimmer will be dominated by propulsion along its axis of symmetry, i.e.~$U \skew3\hat{\zzb}$. In this case, an effective chemotactic swimmer must be able to align with the chemical gradient and then propel along its axis, which corresponds to non-zero values for both $ U_{\text{Janus}}$ and $\Phi_{\text{Janus}}$. In this situation, it is crucial that the two hemispheres of the Janus sphere have different surface activities as well as different surface mobilities. If $ U_{\text{Janus}}$ and $\Phi_{\text{Janus}}$ have the same sign then the swimmer will perform positive chemotaxis, whereas if they have opposite signs the swimmer will perform negative chemotaxis. For intermediate values of the chemical gradient, the optimal strategy for chemotaxis will consist of a combination of the two mechanisms. 

\section{Long-time behaviour: Artificial chemotaxis of a population of non-interacting particles} \label{section-3}

\subsection{Motivation}
Having established the physical law which describes the instantaneous movement of an individual swimmer, we now want to develop an understanding of the long-time behaviour of a phoretic swimmer, or equivalently that of a large number of non-interacting swimmers. Although we neglect the interaction between particles, we focus on understanding the effects of another essential ingredient - stochasticity. Thermal noise plays an important role on the scale of   phoretic swimmers, which are typically a few microns in diameter. The orientation of the swimmers will fluctuate as a result of thermal energy in the surrounding fluid, leading to  diffusive behaviour characterised by a rotational diffusivity $D_r$ given by the Einstein-Smoluchowski relation \citep{Einstein1905,Smolu1906}
\begin{equation}
D_r = \frac{k_B T}{8\pi\eta R^3},
\end{equation}
where $k_B$ is  Boltzmann's constant, $T$ is the temperature,    
 and $\eta$ is the dynamic viscosity of the fluid.  

The physical basis for a macroscopic continuum model lies in the competition between the stochastic effect of rotational diffusion and the deterministic processes by which the swimmer propels and  aligns with the external chemical gradient. If we consider the system on a timescale much larger than the characteristic timescale for rotational diffusion, such that we are essentially averaging over fluctuations in the swimmer orientation, we would expect to see the spatial distribution of swimmers evolve in time according to an effective advection-diffusion equation. The purpose of this section is to quantify this process rigorously. 

The first modelling approach we considered was based on the classical continuum model for gyrotactic swimming micro-organisms proposed by \citet{Pedley90}. This is a  simple and intuitive model, but its simplicity comes at a cost. The authors use a phenomenological definition of the diffusivity tensor which requires the introduction of a direction correlation time, also called relaxation time by some authors \citep[see][]{Batchelor76}. This quantity is not calculated explicitly in their paper, and would have to be estimated from experiments or numerical simulations. Furthermore, the authors assume that the direction correlation time is isotropic, an  assumption  probably not suitable for a swimmer that has a preferential orientation. 

Therefore, we turned instead our attention to the theory of generalised Taylor dispersion (GTD) that Frankel and Brenner used for the study of orientable Brownian particles \citep{Frank91, Frank93} and which has been successfully applied to gyrotactic swimming micro-organisms in more recent years \citep[][]{Hill02, Frank03, Bees12}. By addressing the coupling between dynamics in the orientational space and in the physical space, GTD theory provides a rigorous approach to modelling collections of orientable particles. 

Fundamentally, generalised Taylor dispersion theory uses a moment expansion for the probability density of swimmers, in a similar way to \citet{golestanian2012,pohl2014,bickel2014}. What is novel in the present paper is that the chemical reactions considered in Section 2 lead to a more complex instantaneous behaviour than previously investigated, which brings a further degree of freedom to our continuum model (what we later call the ``indirect chemotactic index") and new phenomenology along with it. Furthermore, the diffusivity tensor is isotropic in all of the above references, because none of them possess the necessary combination of active reorientation and orientation-dependent velocity which promotes the emergence of anisotropic diffusion, as is the case with the phoretic swimmers in our paper.

\subsection{Setup}

We consider a uniform substrate gradient, as in the derivation of the instantaneous behaviour, and no background flow. Since the orientation of the swimmer changes in time, it no longer makes sense to define one of the principal Cartesian axes along the axis of symmetry of the swimmer. Instead, we take the positive $z$-axis to be aligned with the fixed chemical gradient such that $\bnabla s^\infty = |\bnabla s^\infty| \kkb$. We denote the  position of the swimmer by $\xxb$ and its axis of symmetry using the unit vector $\nnb = (\sin\theta \cos\phi, \sin\theta\sin\phi, \cos\theta)$, where $\theta$ and $\phi$ are the usual polar and azimuthal angles from spherical coordinates. 

In this notation, the linear velocity of the phoretic swimmer in \eqref{summary-linear}  can be written as
\begin{equation}
\VVb(\xxb,\nnb) = U(\xxb)\big ( \nnb + \nu(\xxb) \kkb + \mu(\xxb) \nnb\nnb\bcdot\kkb \big ),
\label{linear-velocity}
\end{equation}
where we have introduced two dimensionless parameters
\begin{equation}
\nu(\xxb) \equiv \frac{\alpha |\bnabla s^\infty|}{U(\xxb)} \quad \text{and} \quad \mu(\xxb) \equiv \frac{\beta |\bnabla s^\infty|}{U(\xxb)}\cdot
\end{equation} 
We call $\nu$ and $\mu$ the direct and indirect chemotactic indices, respectively, since they measure the relative effectiveness of direct chemotactic sedimentation, $\alpha \bnabla s^\infty$, and indirect chemotactic sedimentation, $\beta \nnb\nnb\bcdot\bnabla s^\infty$, against active propulsion, $U(\xxb)\nnb$. 

Note that the   relative magnitude $\nu/\mu$ of direct to indirect chemotactic sedimentation determines the angle at which the phoretic swimmer sediments relative to the chemical gradient. This is similar to the Stokes flow sedimentation of an elongated object, such as a rod, under the action of gravity. In Stokes flow the sedimentation angle is given by the geometry (and the drag coefficients) of the elongated body, while in our case it is a function of the chemical properties of the spherical swimmer.

Lastly, we need to consider the mechanism by which the swimmer aligns with the chemical gradient. The orientation of the swimmer evolves in time according to 
\begin{equation}
\dot{\nnb} = \omb \wedge \nnb,
\end{equation}
and using \eqref{summary-angular} we arrive at the following reorientation law
\begin{equation}
\dot{\nnb} = \Omega [\kkb - (\kkb\bcdot\nnb) \ \nnb], 
\label{reorient-law}
\end{equation}
where $\Omega = \Phi |\bnabla s^\infty|$. This prompts us to define a third dimensionless parameter
\begin{equation}\label{RPe}
\lambda\equiv \frac{\Omega}{D_r},
\end{equation}
which is a rotational P\'{e}clet number measuring the relative importance of active reorientation to rotational diffusion. 
 
\subsection{Generalised Taylor dispersion theory}

Although generalised Taylor dispersion theory is now a classical tool, it is useful to clearly introduce it and define the differential operator involved in it. We start from the conservation equation for $P(\nnb,\xxb,t)$, the probability density function of finding a swimmer with orientation $\nnb$ at position $\xxb$ at time $t$, which is satisfies the conservation equation
\begin{equation}
\frac{\p P}{\p t} + \bnabla_r \bcdot \big (\VVb(\xxb,\nnb) P - D_t \bnabla_r P \big )  + \bnabla_n \bcdot \big (\dot{\nnb} P - D_r \bnabla_n P \big ) = 0,
\label{eq-full-cons}
\end{equation}
where the differential operators for physical space and orientational space are 
\begin{equation}
\bnabla_r \equiv \boldsymbol{i} \frac{\p}{\p x_1} + \boldsymbol{j} \frac{\p}{\p x_2} + \boldsymbol{k} \frac{\p}{\p x_3}, \quad \bnabla_n \equiv \skew3\hat{\thb} \frac{\p}{\p \theta} + \skew3\hat{\phb} \frac{1}{\sin\theta} \frac{\p}{\p \phi},
\end{equation}
respectively. The law in \eqref{eq-full-cons} says that the local rate of change in the distribution of swimmers is due to the flux of swimmers in physical space (advection and translational diffusion) and the flux of swimmers in orientational space (advection and rotational diffusion).

Since we are  interested in the spatial distribution of the particles,   we define the particle density
\begin{equation}
\rho(\xxb,t) \equiv \int_{S_2} P(\nnb,\xxb,t)~\mathrm{d}^2\nnb.
\end{equation}
Upon careful consideration of the moments of $P(\nnb,\xxb,t)$, Frankel and Brenner showed that at the macroscopic level, and on timescales $t \gg D_r^{-1}$, a population of non-interacting orientable particles obeys the  advection-diffusion equation
\begin{equation}
\frac{\p \rho}{\p t} + \bnabla_{\rrb}\bcdot\big (\VVb_s \rho - \boldsymbol{D} \bcdot \bnabla_{\rrb}\rho\big ) = 0
\label{eq:gtdmacro},
\end{equation}
where $\VVb_s$ is the mean swimming velocity and $\boldsymbol{D}$ is the diffusivity tensor. The mean swimming velocity is given by
\begin{equation}
\VVb_s (\xxb) = \int_{S_2}f(\nnb) \VVb(\xxb,\nnb)~ \mathrm{d}^2\nnb,
\label{eq-GTDT-Vs-defn}
\end{equation}
whereas the diffusivity tensor is the sum of contributions from passive translational diffusion ($\boldsymbol{D}_{\text{pass}}$) and active phoretic motion ($\boldsymbol{D}_{\text{act}}$) as
\begin{equation}
\boldsymbol{D}(\xxb) = \boldsymbol{D}_{\text{pass}} + \boldsymbol{D}_{\text{act}}(\xxb).
\end{equation}

The passive part of the diffusivity tensor, $\boldsymbol{D}_{\text{pass}}$, is simply the average of the translational diffusivity tensor, $\boldsymbol{D}_t(\nnb)$, over the space of orientations \citep[see][]{Frank93}. For a spherical swimmer it is reasonable to assume that Brownian noise acts isotropically, and so $\boldsymbol{D}_{\text{pass}} = D_t \boldsymbol{I}$ with $D_t$ being the translational diffusion coefficient. Recently it has been shown by \citet{Agudo2018} that, at the nano scale, spatial variations in the translational diffusion coefficient of enzymes can compete with phoretic effects, and it may well be possible that the catalytic coating of phoretic swimmers could lead to enchanced diffusion coefficients as the swimmer moves through the chemical gradient. However, due to a lack of documentation on the enhanced diffusion capabilities of a microscopic phoretic swimmer, and due to the limitations of generalised Taylor dispersion theory, to consider spatial variations of $D_t$ would go beyond the scope of this paper. 

With the assumption that $\boldsymbol{D}_{\text{pass}}$ is constant and independent of the phoretic mechanisms for motion, we will focus from here onwards on quantifying the active diffusivity tensor given by 
\begin{equation}
\boldsymbol{D}_{\text{act}}(\xxb) = \int_{S_2} [\bbb(\xxb,\nnb) \VVb(\xxb,\nnb)]^{\text{sym}} \mathrm{d}^2\nnb,
\label{eq-GTDT-D-defn}
\end{equation}
where $[\dots ]^{\text{sym}}$ denotes the symmetric part of the tensor. 

The expressions for $\VVb_s (\xxb)$ and $\boldsymbol{D}_{\text{act}}(\xxb)$ contain two new fields that have to be computed:  $f(\nnb)$ is the long-time steady distribution of the particle orientation, and $\bbb(\xxb,\nnb)$ represents the long-time relative displacement of a particle given that its instantaneous orientation is $\nnb$, compared to its expected position over all possible orientations. For a formal definition of these quantities we need introduce the linear operator
\begin{equation}
\Lin (\star) = \bnabla_{\nnb} \bcdot \left[ \dot{\nnb} (\star) - D_r \bnabla_{\nnb} (\star) \right],
\end{equation}
which describes the flux of a quantity in orientational space. Then $f$ and $\bbb$ are the solutions to the following equations and boundary conditions
\begin{eqnarray}
\Lin f &=& 0, \quad \int_{S_2}f(\nnb) \mathrm{d}^2\nnb = 1,    \label{eq:f}\\
\Lin \bbb &=& f(\nnb) (\VVb(\xxb,\nnb) - \VVb_s(\xxb)), \quad \int_{S_2} \bbb ~\mathrm{d}^2\nnb = 0
\label{eq:b}
\end{eqnarray}
For a full derivation of these results see \citet{Frank93} or, for a more succinct account,   \citet{Hill02}. 

To provide some physical intuition for these two equations, we observe that equation (\ref{eq:f}) corresponds to solving $\p f/\p t =0$ and therefore finding the steady-state distribution of orientations subject to a global normalisation condition. Similarly, equation (\ref{eq:b}) can be thought of as solving $\p \bbb/\p t =f(\nnb) (\VVb(\xxb,\nnb) - \VVb_s(\xxb))$ which means that the rate of change of the relative displacement is due to the relative velocity, weighted by the probability of finding a swimmer with that specific orientation. The obvious boundary condition to impose here is that, averaged over all possible orientations, the relative displacement must vanish. 

One important difference between our application of GTD theory to phoretic swimmers compared to previous applications to orientable Brownian particles and gyrotactic micro-swimmers is that the swimming velocity $\VVb(\xxb,\nnb)$ of phoretic swimmers depends on position as well. The classical theory still applies provided that the velocity is slowly-varying in space, such that the process of averaging over particle orientations can be carried out on timescales much larger than the diffusive timescale, $D_r^{-1}$, but much smaller than the timescale of travel to a region of significantly different swimming speed. So there must exist a timescale $\tau$, satisfying
\begin{equation}
D_r^{-1} \ll \tau \ll |\nabla U|^{-1},
\end{equation}
on which we can average \eqref{eq-full-cons} over the space of swimmer orientations. In particular, we must have
\begin{equation}
|\nabla U| \ll D_r.
\end{equation}
In that case, the advection-diffusion equation \eqref{eq:gtdmacro} will be valid on timescales greater than or equal to $\tau$, and will involve a mean swimming velocity, $\VVb_s (\xxb)$,  and a diffusivity tensor, $\boldsymbol{D}(\xxb)$, which are slowly-varying in space. Under this  approximation, quantities such as $\nu(\xxb), \mu(\xxb), \VVb(\xxb,\nnb)$ and $\bbb(\xxb,\nnb)$ are also slowly-varying in space, and we will drop the explicit mention of spatial dependence from our notation.

\subsection{Distribution of swimmer orientations}
Since the orientation of the swimmers fluctuates over time from thermal noise, the only fixed meaningful direction in our problem is that of the chemical gradient. Therefore, quantities such as $f(\nnb)$ and $\bbb(\nnb)$ will inherit rotational symmetry around the direction of the chemical gradient (the $\kkb$ axis). For such an axisymmetric problem, the linear operator simplifies to
\begin{equation}
    \Lin (\star) = -\frac{\Omega}{\sin\theta}\frac{\p}{\p\theta}\left(\sin^2\theta(\star)\right) - \frac{D_r}{\sin\theta}\frac{\p}{\p\theta}\left(\sin\theta\frac{\p(\star)}{\p\theta}\right).
\end{equation}
Then   \eqref{eq:f} has a trivial first integral 
\begin{equation}
 \lambda\sin^2\theta f + \sin\theta \frac{\p f}{\p\theta}= 0 ,
\end{equation}
where the integrating constant can be set to zero assuming that $f$ is well-behaved at the poles. The solution to this, subject to the appropriate normalisation condition, is the same as that found by \citet{Pedley90}. The long-time steady distribution of the swimmer orientation is 
\begin{equation}
    f(\nnb) = \hat{f}(\cos\theta) = \frac{\lambda e^{\lambda \cos\theta}}{4\upi \sinh\lambda},
    \label{eq-equilibrium-orientation}
\end{equation}
where $\lambda$ is the rotational P\'{e}clet number, \eqref{RPe}.

\subsection{Mean swimming velocity}
From equations \eqref{linear-velocity} and \eqref{eq-GTDT-Vs-defn}, the mean swimming velocity  is obtained as
\begin{equation}
\VVb_s = U \left( \int_{S_2}f(\nnb)\nnb \ \mathrm{d}^2\nnb + \nu \int_{S_2}f(\nnb)\kkb \ \mathrm{d}^2\nnb + \mu \int_{S_2} f(\nnb)\nnb\nnb\bcdot\kkb \ \mathrm{d}^2\nnb  \right).
\end{equation}
We use the normalisation of $f(\nnb)$ to simplify the second integral, and   average over the azimuthal angle in the other two integrals in order to get
\begin{equation}
\VVb_s = U \left( \int_0^\pi \hat{f}(\cos\theta)\cos\theta \sin\theta \mathrm{d}\theta + \nu + \mu \int_0^\pi \hat{f}(\cos\theta)\cos^2\theta \sin\theta \mathrm{d}\theta  \right) \kkb.
\end{equation}
Using our previous expression for $\hat{f}(\cos\theta)$, we obtain the following analytical expression for the mean swimming velocity
\begin{equation}
\VVb_s = U \left[\frac{-1 + \lambda \coth\lambda}{\lambda} + \nu + \mu \left(\frac{2+\lambda^2-2\lambda\coth\lambda}{\lambda^2}\right)\right] \kkb.
\end{equation}

Note that the linear velocity of the swimmer relative to the mean swimming velocity is a function of $\lambda$ and $\mu$ only since
\begin{equation}
\VVb(\nnb) - \VVb_s = U \left[ \nnb + \mu \nnb\nnb\bcdot\kkb - \tilde{u}(\lambda,\mu) \kkb\right]
\label{eq:relative_velocity},
\end{equation}
where we have defined for convenience
\begin{equation}
\tilde{u}(\lambda,\mu) \equiv \frac{-1 + \lambda \coth\lambda}{\lambda} + \mu \left(\frac{2+\lambda^2-2\lambda\coth\lambda}{\lambda^2}\right).
\end{equation}
As a result,  both $\bbb(\nnb)$ and the diffusivity tensor $\boldsymbol{D}_{\text{act}}$ will then only depend on the value of  $\lambda$ and $\mu$. Physically,  the direct chemotactic index $\nu$ corresponds to a component of velocity that is constant in time which  imparts a constant drift to the phoretic swimmer without affecting the way in which the swimmer diffuses through space.

\subsection{Active diffusivity tensor}
The first step in calculating the active diffusivity tensor is to solve for the vector field $\bbb(\nnb)$. The expression for the relative swimming velocity from   \eqref{eq:relative_velocity} together with the definition of $\bbb(\nnb)$ from equation (\ref{eq:b}) suggest that the only dependence on the azimuthal angle that can be excited in  $\bbb(\nnb)$ is a $\cos\phi$ dependence in its $x$-component and a $\sin\phi$ dependence in its $y$-component. Therefore, we look to solve for $\bbb(\nnb)$  in the form 
\begin{equation}
\bbb(\nnb) = (b_\perp(\theta)\cos\phi,b_\perp(\theta)\sin\phi,b_\parallel(\theta)),
\end{equation}
where the function $b_\perp(\theta)$ is the same in the $x$ and $y$-components due to rotational symmetry around the direction of the chemical gradient. 
We then decompose
\begin{eqnarray}
b_\perp(\theta) &=& \frac{U}{D_r}\sum_{l=1}^{\infty} - b^\perp_l P_l^1(\cos\theta) = \frac{U}{D_r}\sum_{l=1}^{\infty} b^\perp_l \sin\theta P_l'(\cos\theta),\\
b_\parallel(\theta) &=& \frac{U}{D_r}\sum_{l=0}^{\infty} b^\parallel_l P_l(\cos\theta),
\end{eqnarray}
and substitute   into equation (\ref{eq:b}) to obtain, after some simplification, the equalities
\begin{eqnarray}
\sum_{l=1}^{\infty} b^\perp_l \left[\left(l(l+1)-3\lambda\cos\theta\right) P_l' + \lambda\sin^2\theta P_l''\right] &=& (1+\mu\cos\theta) \hat{f}(\cos\theta), \label{eq:bperp}\\
\sum_{l=1}^{\infty} b^\parallel_l \left[\left(l(l+1)-2\lambda\cos\theta\right) P_l + \lambda\sin^2\theta P_l'\right] &=& \left(\cos\theta(1+\mu\cos\theta)-\tilde{u}\right) \hat{f}(\cos\theta) \label{eq:bparl}.\quad
\end{eqnarray}
Note that $b^\parallel_0 = 0$ results from imposing $\int_{S_2} \bbb~d^2\nnb = 0$.

The most important step in the derivation is choosing the appropriate inner product for these two equations, so that the integrals on the left-hand side may be evaluated exactly. In equation (\ref{eq:bperp}) we make the substitution $\xi = \cos\theta$ and take $\int \dots (1-\xi^2) P_m(\xi) d\xi$ of both sides. The left-hand side can then be evaluated using the identities \eqref{eq-1-u2-Pk'-Pl}, \eqref{eq-(1-u2)uPk_Pl'} and \eqref{eq-(1-u2)2_Pk_Pl''}. We obtain a set of linear equations for the coefficients $b_l^\perp$ as 
\begin{equation}
P_{ml}b_l^\perp = \int_{-1}^{+1} (1+\mu\xi) \hat{f}(\xi) (1-\xi^2) P_m(\xi) \mathrm{d}\xi, \quad (m\geq 0) 
\label{eq:pentadiagonal}
\end{equation}
where the pentadiagonal matrix $\boldsymbol{P}$ has components
\begin{multline}
P_{ml} = \frac{2\lambda(m+3)(m+2)(m+1)^2}{(2m+5)(2m+3)(2m+1)} \delta_{l,m+2} \\ 
+ \frac{2(m+2)^2(m+1)^2}{(2m+3)(2m+1)} \delta_{l,m+1} \qquad \qquad \qquad \qquad  \\ 
- \frac{2\lambda m(m+1)(2m^2+2m-1)}{(2m+3)(2m+1)(2m-1)} \delta_{lm} \\ 
\qquad \qquad \qquad + \frac{2m^2(m-1)^2}{(2m+1)(2m-1)} \delta_{l,m-1} \\ 
+ \frac{2\lambda m^2(m-1)(m-2)}{(2m+1)(2m-1)(2m-3)} \delta_{l,m-2}.
\end{multline}

In equation (\ref{eq:bparl}) we make the same substitution and take $\int \dots P_m(\xi) d\xi$ of both sides. Using identities \eqref{eq-Legendre-ortho-cond}, \eqref{eq-uPkPl-integral} and \eqref{eq-1-u2-Pk'-Pl} we obtain another set of linear equations, this time for the coefficients $b_l^\parallel$
\begin{equation}
T_{ml}b_l^\parallel = \int_{-1}^{+1} \left[\xi(1+\mu\xi) - \tilde{u}\right] \hat{f}(\xi) P_m(\xi) \mathrm{d}\xi, \quad (m\geq 0)
\label{eq:tridiagonal}
\end{equation}
where the tridiagonal matrix $\boldsymbol{T}$ has components
\begin{multline}
\qquad \qquad \qquad T_{ml} = \frac{2\lambda m(m+1)}{(2m+1)(2m+3)} \delta_{l,m+1} \\
+ \frac{2m(m+1)}{2m+1} \delta_{lm}\\
- \frac{2\lambda m(m+1)}{(2m-1)(2m+1)} \delta_{l,m-1}. \qquad \qquad
\label{eq:tridiagonal-components}
\end{multline}

Equations \eqref{eq:pentadiagonal} and \eqref{eq:tridiagonal} represent infinite systems of linear equations for the coefficients $b_{m}^\perp$ and $b_{m}^\parallel$ which cannot be inverted analytically. Progress can be made by truncating the series for $b_\perp(\theta)$ and $b_\parallel(\theta)$ to $N$ terms and inverting the resulting $N$-by-$N$ systems numerically. Rather fortunately, only the first two coefficients are needed to compute the active diffusivity tensor, as  we shall now see.

If we substitute our expressions for $\VVb(\nnb)$ and $\bbb(\nnb)$ into the definition of the active diffusivity tensor, \eqref{eq-GTDT-D-defn}, and average over the azimuthal angle, we find that all the off-diagonal terms vanish. Furthermore, we have $(\boldsymbol{D}_{\text{act}})_{11} = (\boldsymbol{D}_{\text{act}})_{22}$ due to rotational symmetry about the $\kkb$ axis, which is the direction of the chemical gradient. We denote $D_\parallel\equiv (\boldsymbol{D}_{\text{act}})_{33}$ to be the active diffusivity parallel to the chemical gradient, and $D_\perp\equiv (\boldsymbol{D}_{\text{act}})_{11} = (\boldsymbol{D}_{\text{act}})_{22}$ to be the active diffusivity in the plane normal to it. From  \eqref{eq-GTDT-D-defn} these quantities are   
\begin{eqnarray}
D_\perp &=& ~\upi \int_0^\pi b_\perp(\theta) \ U\sin\theta(1+\mu\cos\theta) \sin\theta \mathrm{d}\theta, \\ D_\parallel &=& 2\upi \int_0^\pi b_\parallel(\theta) \ U\cos\theta(1+\mu\cos\theta) \sin\theta \mathrm{d}\theta.
\end{eqnarray}
Using identities \eqref{eq-A-firstfew}, \eqref{eq-uPkPl-integral}, \eqref{eq-1-u2-Pk'-Pl} we can simplify these expression to  finally obtain
\begin{eqnarray}
D_\perp &=& \frac{4\pi U^2}{D_r}\left(\frac{b_1^\perp}{3}+\frac{\mu b_2^\perp}{5}\right), 
\label{eq:Dperp_result}\\ D_\parallel &=& \frac{4\pi U^2}{D_r}\left(\frac{b_1^\parallel}{3}+\frac{2\mu b_2^\parallel}{15}\right).
\label{eq:Dparl_result}
\end{eqnarray}
Since only the first two coefficients $b_{m}^\perp$ and $b_{m}^\parallel$  ($m=1,2$) are needed in the calculation of the diffusivity tensor, the truncated systems converge very rapidly and we find that it is usually sufficient to truncate to $N$ to order $\mathcal{O}(\lambda,\mu)$ or $N=2$, whichever is larger.

\subsection{Numerical simulations}

\begin{figure}
	\begin{minipage}[b]{0.5\textwidth}
		\centering
		\includegraphics[width=\textwidth]{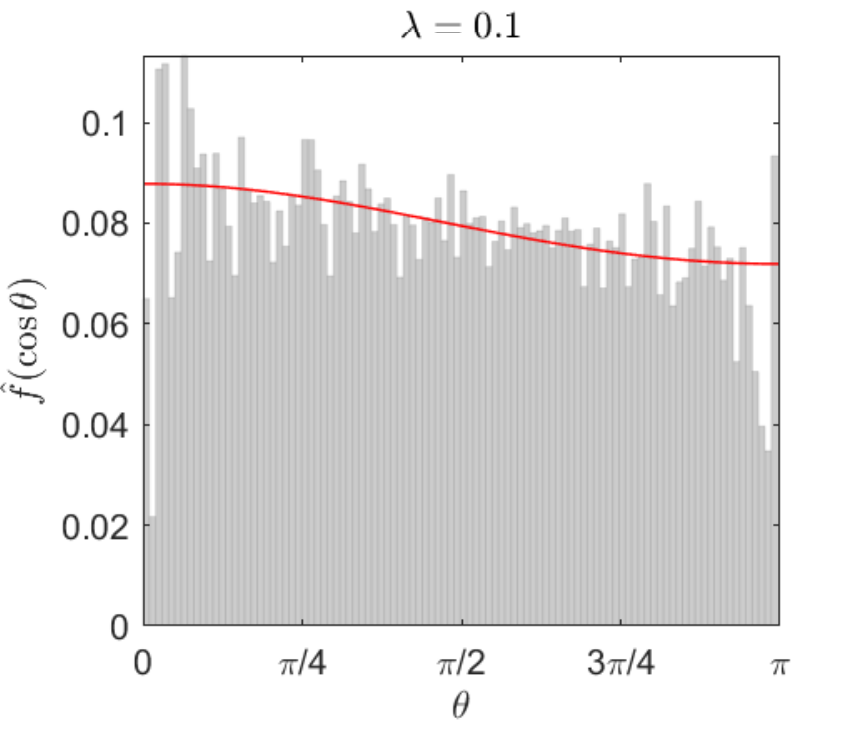}
	\end{minipage}
	\begin{minipage}[b]{0.5\textwidth}
		\centering
		\includegraphics[width=\textwidth]{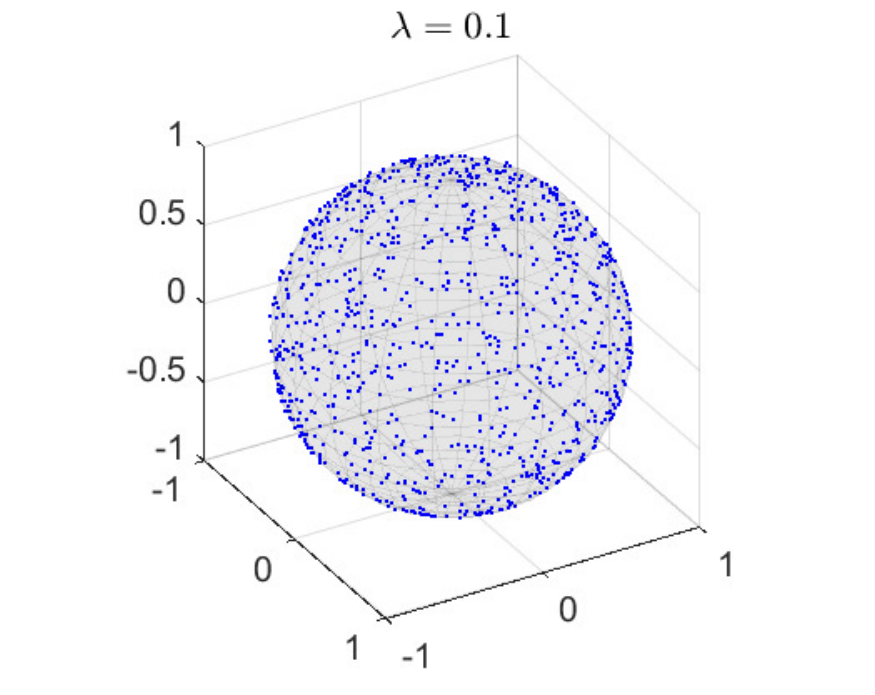}
	\end{minipage}
	\begin{minipage}[b]{0.5\textwidth}
		\centering
		\includegraphics[width=\textwidth]{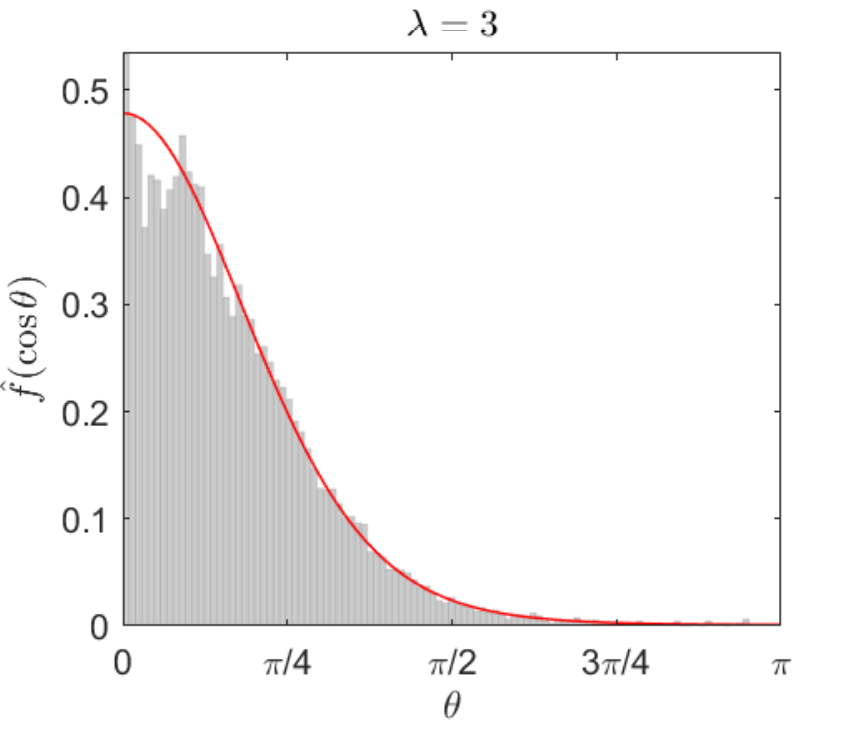}
	\end{minipage}
	\begin{minipage}[b]{0.5\textwidth}
		\centering
		\includegraphics[width=\textwidth]{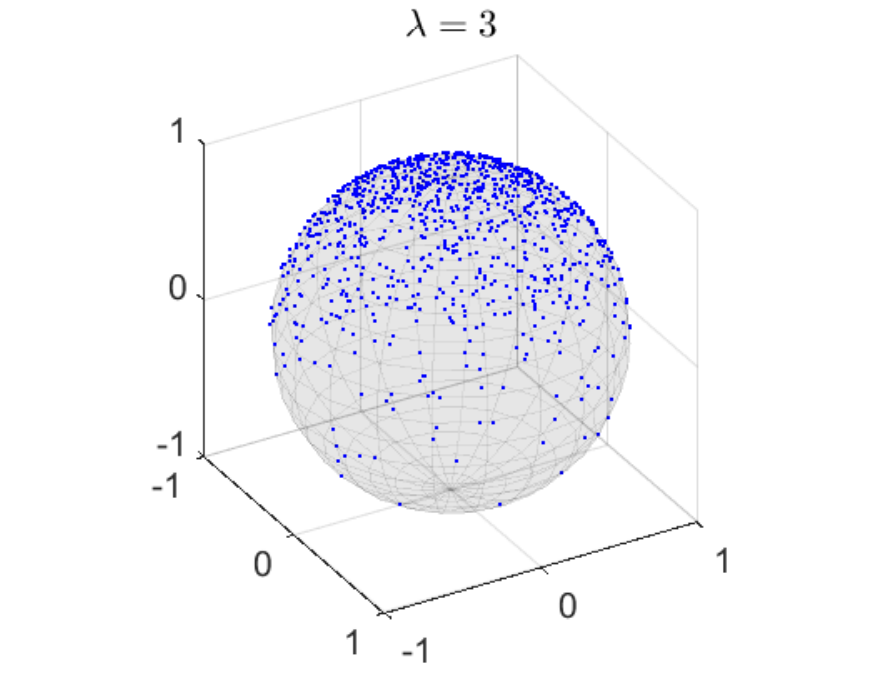}
	\end{minipage}
	\begin{minipage}[b]{0.5\textwidth}
		\centering
		\includegraphics[width=\textwidth]{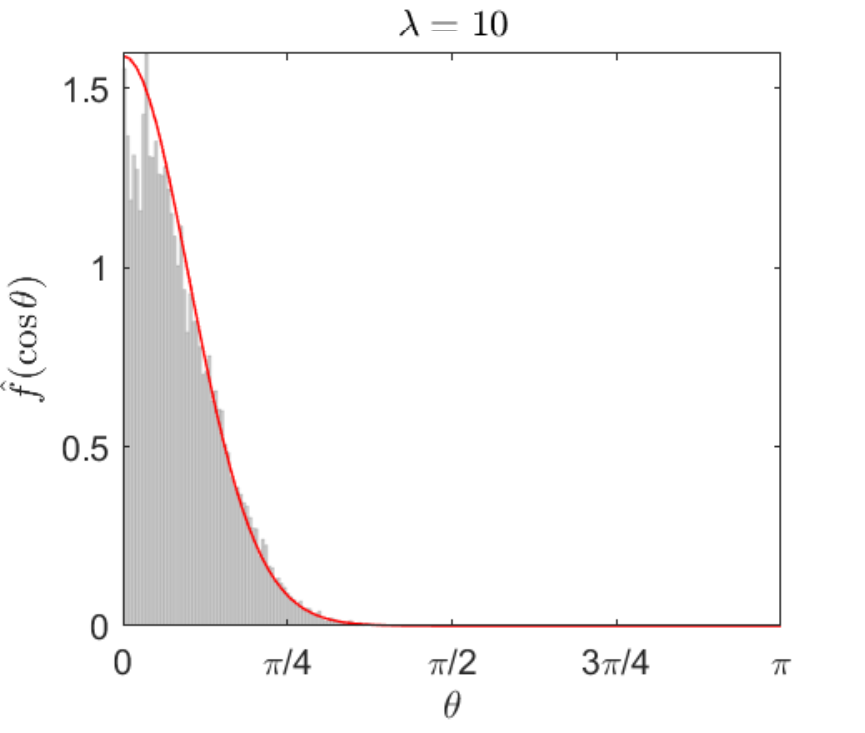}
	\end{minipage}
	\begin{minipage}[b]{0.5\textwidth}
		\centering
		\includegraphics[width=\textwidth]{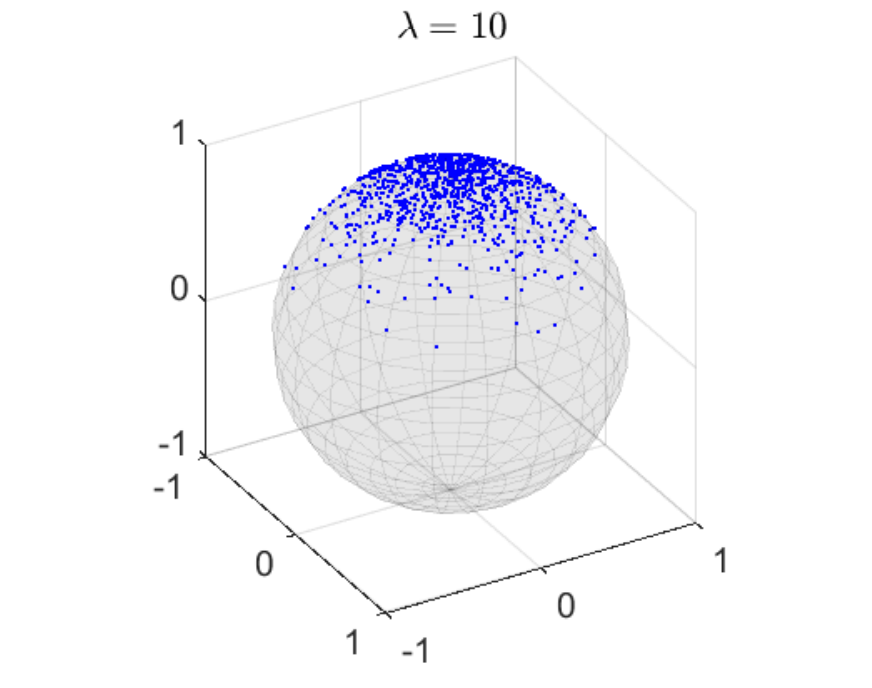}
	\end{minipage}
	\caption{Comparison of the distribution of swimmer orientations obtained from numerical simulations with the predictions of the continuum model. Left:  histograms of final orientations for 10,000 swimmers whose initial orientations were chosen from a uniform distribution, after being evolved in time for 500 time steps of size $\Delta t = 10^{-2} D_r^{-1}$. The red solid line corresponds to the analytical expression from   \eqref{eq-equilibrium-orientation} for the steady-state distribution of swimmer orientations. Right: sample of 1,000 swimmer orientations chosen from the steady-state distribution. Each swimmer orientation corresponds to a unit vector which is represented here as a blue dot on the unit sphere. From top to bottom the rotational P\'{e}clet number is $\lambda = 0.1, 3, 10$.}
	\label{fig:orientation}
\end{figure}

In order to validate the continuum model we performed numerical simulations of phoretic swimmers using a simple Euler-Maruyama scheme. To capture the correct dynamics we use a time step much smaller than the timescale $D_r^{-1}$ for rotational diffusion, and calculate all relevant macroscopic quantities as averages over a time period larger than $D_r^{-1}$. At each time step, the position and orientation of the swimmers are updated with the deterministic contribution from the linear and angular velocities of the swimmer, as in the classical Euler method. In addition to this, the orientation of the swimmer experiences a stochastic contribution due to thermal noise. During the $n$th time step the axis of symmetry of the swimmer is deflected in a random direction by an angle $\Delta \theta_n = 2 \sqrt{D_r \Delta t} \Delta W_n$, where $\Delta W_n$ are independent random variables taken from the standard normal distribution \citep[][]{saragosti12}. 

The only parameters that enter the simulation are the phenomenological constants $\mu$ and $\lambda$ which describe deterministic translation and reorientation, while the stochastic reorientation timescale $D_r^{-1}$ is normalised to one unit of time in our simulations. Since the direct chemotactic index $\nu$  contributes to a constant linear velocity of the particle, we can take it to be zero in our simulations, which means that we are effectively moving to a frame of reference that is sedimenting with the particle, translating at a constant speed $\alpha |\bnabla s^\infty|$ in the direction of the chemical gradient. We only consider a non-zero value for $\nu$ in figure \ref{fig:diffusivities} when we normalise the active diffusivity coefficients (derived from theory only, no simulations) using the mean swimming velocity. In this case we take $\nu$ to be an arbitrary non-zero value in order to avoid the singularity at $\lambda, \mu = 0$ where the mean swimming velocity would also go to zero otherwise.

To validate our numerical method we first compared the steady-state distribution of orientations for our simulated swimmers against the analytical expression in   \eqref{eq-equilibrium-orientation}, which is a well-established result in the literature \citep[][]{Pedley90,Bees12}. This validation is illustrated in  figure~\ref{fig:orientation} which shows    perfect agreement between numerical  simulations and theory.

\begin{figure}
	\centering
	\includegraphics[width=\textwidth]{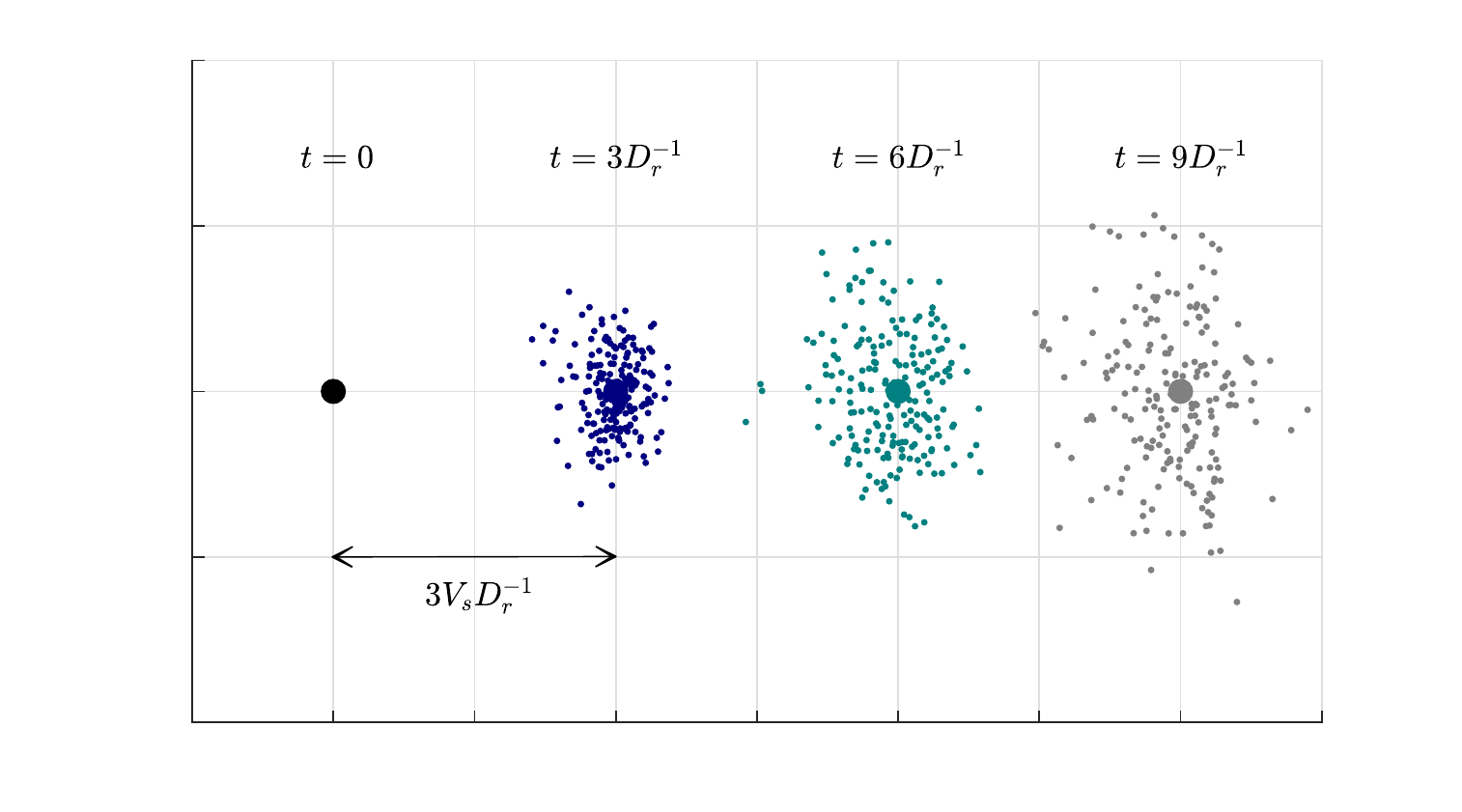}
	\caption{Time evolution of 200 independent trajectories, all starting from the origin at $t=0$. Each snapshot is taken after a time interval of $3D_r^{-1}$, and the chemical gradient is along the horizontal axis. The large dots are placed at regular distances of $3V_s D_r^{-1}$ from the origin which confirms that the cloud of phoretic swimmers drifts along the direction of the chemical gradient at the mean swimming velocity ($V_s $), as predicted by the continuum model. This simulation corresponds to parameters $\lambda=5$ and $\mu=3$, and the time step used was $\Delta t = 10^{-2} D_r^{-1}$.}
	\label{fig:cloud}
\end{figure}

We then investigated qualitatively the behaviour of a cloud of swimmers initially starting from the origin. These results are   depicted in figure~\ref{fig:cloud}. The cloud of swimmers is seen to diffuse anisotropically and to drift along the direction of the chemical gradient at the mean swimming velocity, $V_s$,  as predicted by the continuum model.

\begin{figure}
	\begin{minipage}[b]{0.5\textwidth}
		\centering
		\includegraphics[width=\textwidth]{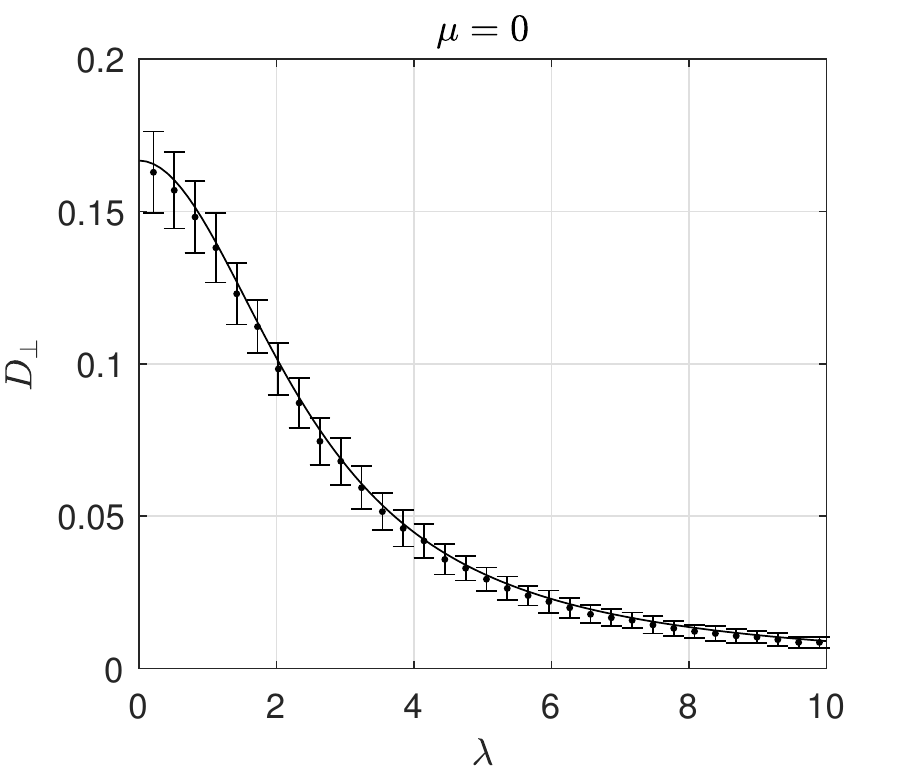}
	\end{minipage}
	\begin{minipage}[b]{0.5\textwidth}
		\centering
		\includegraphics[width=\textwidth]{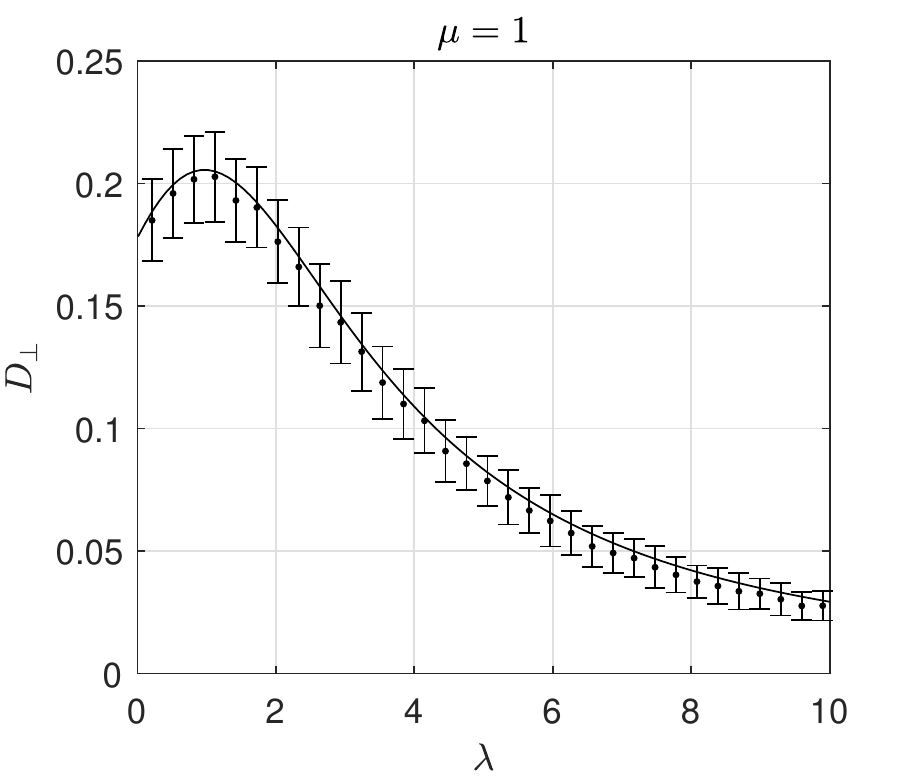}
	\end{minipage}
	\begin{minipage}[b]{0.5\textwidth}
		\centering
		\includegraphics[width=\textwidth]{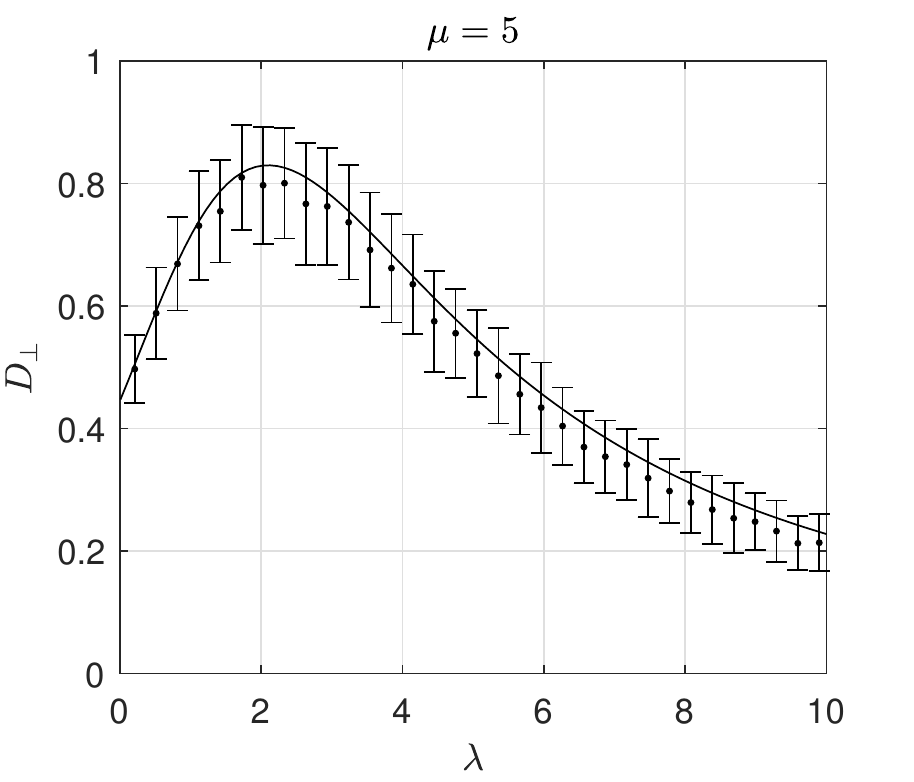}
	\end{minipage}
	\begin{minipage}[b]{0.5\textwidth}
		\centering
		\includegraphics[width=\textwidth]{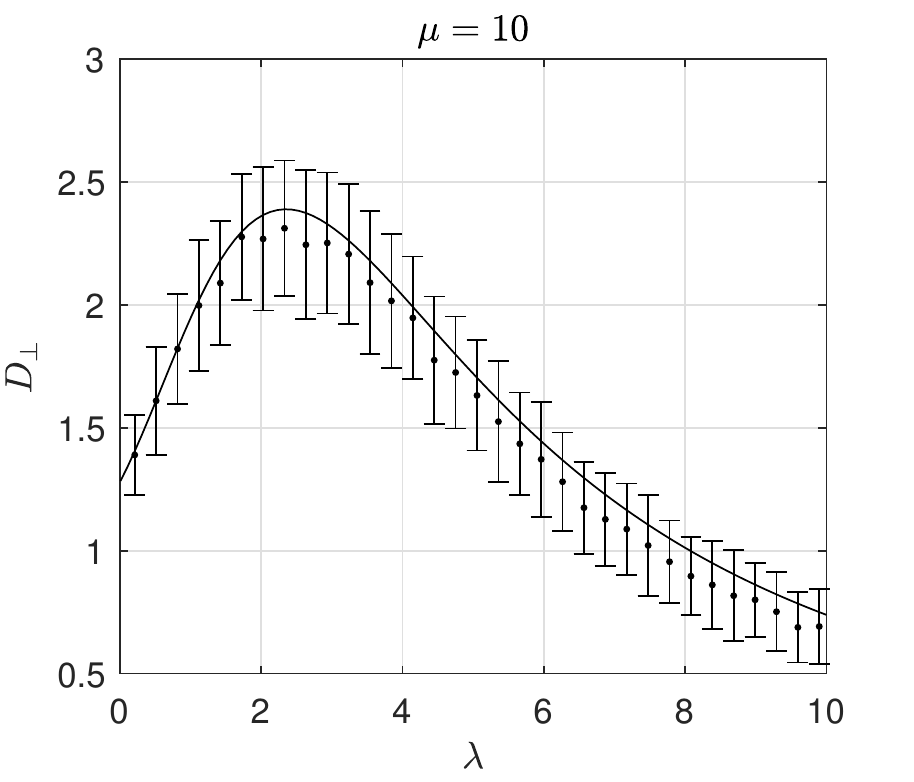}
	\end{minipage}
	\caption{Comparison of the values obtained from numerical simulations against those predicted by generalised Taylor dispersion theory for the dependence of  component $D_\perp$ of the active diffusivity tensor on $\lambda$. The simulations were run with 100 samples of 1,000 swimmers using a time step $\Delta t = 10^{-2} D_r^{-1}$ over a period $2D_r^{-1}$. Each sample was used to calculate one value of the diffusivity $D_\perp$, and then averaged to obtain one data point. The error bars represent one standard deviation amongst the 100 values obtained for the diffusivity. The solid lines represent the predictions of the continuum model from equations \eqref{eq:pentadiagonal}-\eqref{eq:tridiagonal-components} and \eqref{eq:Dperp_result}-\eqref{eq:Dparl_result}.   	In all plots the diffusivity is non-dimensionalised by $U^2D_r^{-1}$. The four graphs correspond to parameter values $\mu =0$, $\mu =1$, $\mu =5$ and $\mu =10$.}
	\label{fig:comparison}
\end{figure}

The final comparison focuses on   the values of the active diffusivity tensor computed from stochastic simulations against those given by generalised Taylor dispersion theory. We plot in figure~\ref{fig:comparison}   four comparative graphs for the dependence  of $D_\perp$ on $\lambda$ at different values of the indirect chemotactic index $\mu$ (results for $D_\parallel$, not shown are very similar).  In all cases, the agreement between the simulations and the theory is excellent and the data points obtained numerically are always within one standard deviation of the theoretical prediction. 

\begin{figure}
	\begin{minipage}[b]{0.5\textwidth}
		\centering
		\includegraphics[width=\textwidth]{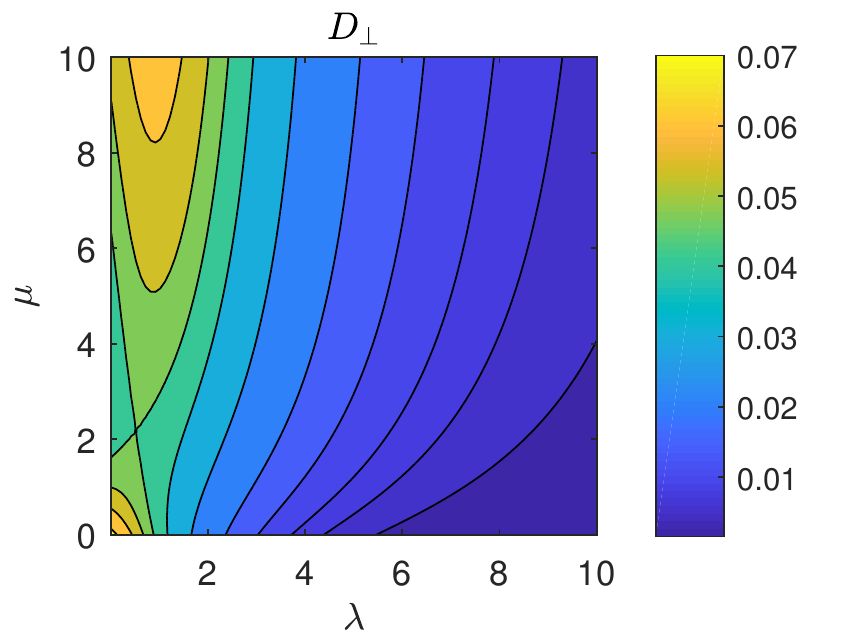}
	\end{minipage}
	\begin{minipage}[b]{0.5\textwidth}
		\centering
		\includegraphics[width=\textwidth]{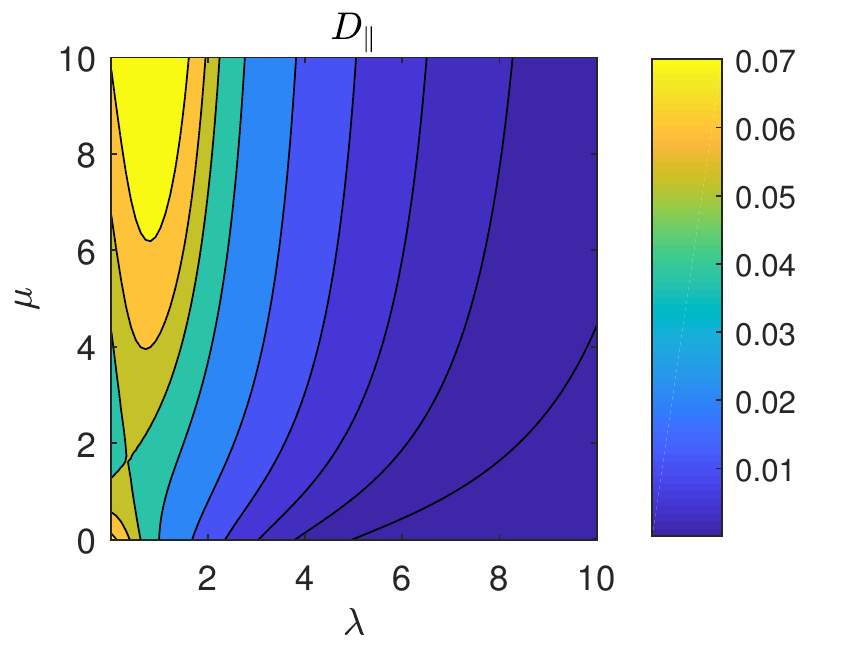}
	\end{minipage}
	\caption{Iso-values of the diffusivities $D_\perp$ and $D_\parallel$ computed using generalised Taylor dispersion theory in the $(\lambda,\mu)$ plane. Both  diffusivities are normalised using the mean swimming velocity and the timescale of rotational diffusion, i.e.~$D \mapsto D/(V_s^2D_r^{-1})$. These contour plots are obtained with a value of $\nu=1.5$ for the direct chemotactic index, but there are no qualitative changes as $\nu$ is varied (not shown). We observe consistent decay of both diffusivities as $\lambda \tti$ at a fixed value of $\mu$, as well as a saddle point in diffusivity for $\lambda,\mu=\mathcal{O}(1)$.}
	\label{fig:diffusivities}
\end{figure}

We can also use the continuum model to understand how the active diffusivity tensor changes as we vary the values of the rotational P\'{e}clet number, $\lambda$, and the indirect chemotactic index, $\mu$. We plot in figure~\ref{fig:diffusivities} iso-values the two components of the active diffusivity tensor normalised by $V_s^2D_r^{-1}$, with $V_s$ being the magnitude of the mean swimming velocity. The active diffusivity decays as the rotational P\'{e}clet number $\lambda \tti$ for any fixed value of $\mu$, as expected, because the active reorientation of the particle enhances directed motion along the chemical gradient and suppresses diffusion. We also observe a local maximum in diffusion at $\lambda,\mu =0$ and a saddle point for $\lambda,\mu$ of $\mathcal{O}(1)$. The surprising finding is that diffusivity increases as $\mu \tti$ for a fixed value of $\lambda$, meaning that indirect chemotactic sedimentation tends to augment diffusion relative to the mean drift velocity of the swimmer.

\begin{figure}
	\centering
	\includegraphics[width=.55\textwidth]{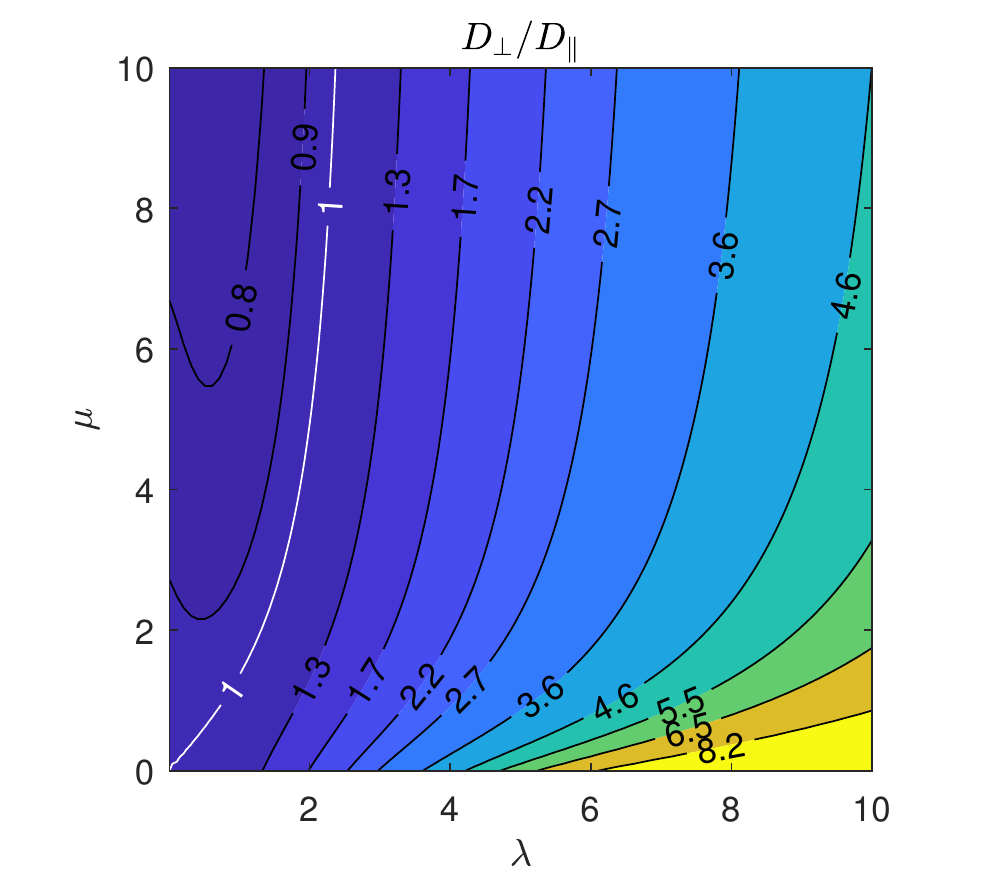}
	\caption{The ratio $D_\perp/D_\parallel$ computed using generalised Taylor dispersion theory in the $(\lambda,\mu)$ plane. As expected, the contour $D_\perp/D_\parallel=1$ passes through the origin, because the system reverts to isotropic diffusion in the limit $\lambda,\mu \ttz$. Increasing $\mu$ above this contour leads to a ratio $D_\perp/D_\parallel < 1$, meaning that indirect chemotactic sedimentation enhances diffusion along the chemical gradient relative to diffusion in the normal plane. On the other hand, increasing $\lambda$ to the right of this contour leads to a ratio $D_\perp/D_\parallel >1$, meaning that chemotactic alignment suppresses diffusion along the chemical gradient relative to diffusion in the normal plane.  }
	\label{fig:ratio}
\end{figure}

Finally, we investigate how the anisotropy of the diffusion depends on the parameters of the problem; specifically we measure the ratio $D_\perp/D_\parallel$ of the diffusivity in the plane   normal to the chemical gradient to the diffusivity parallel to the gradient. Iso-values of this ratio in the $(\lambda,\mu)$ plane are shown in   figure~\ref{fig:ratio}. We find that the contour $D_\perp/D_\parallel = 1$ passes through the origin, in agreement  with our expectation to recover isotropic diffusion in the limit where both chemotactic alignment and indirect chemotactic sedimentation are weak. We observe that the ratio $D_\perp/D_\parallel$ falls below unity in the region of the $(\lambda,\mu)$ plane above this contour and rises above it in the region to the right of the contour. This indicates that chemotactic alignment, whose strength is given by $\lambda$, favours the reduction of diffusion along the chemical gradient. On the other hand, the indirect chemotactic sedimentation quantified by $\mu$ favours the reduction of diffusion in the plane normal to the chemical gradient.

From figure~\ref{fig:ratio} we also note that $D_\perp/D_\parallel > 1$ for the parameters $\lambda=5$ and $\mu=3$ which were used in our analysis of the time evolution of a cloud of swimmers. This is apparent in figure~\ref{fig:cloud} where the cloud of swimmers, initially at the origin, grows anisotropically over time, with more spreading being observed in the vertical than in the horizontal direction, the latter being the direction of the chemical gradient.

\subsection{Asymptotics}

For small $\lambda$ and $\mu$ we can invert the 2-by-2 systems obtained from the truncation of the linear systems of equations \eqref{eq:pentadiagonal} and \eqref{eq:tridiagonal} exactly, in order to obtain the approximate analytical expressions for $D_\perp$ and $D_\parallel$ that are given in Appendix \ref{appC}. We then take the limit $\lambda,\mu \ttz$ in these expressions and find that
\begin{equation}
D_\perp \rightarrow \frac{U^2}{6D_r}, \quad D_\parallel \rightarrow \frac{U^2}{6D_r}
\end{equation}
which is precisely what we would expect in the purely diffusive limit where both chemotactic alignment and indirect chemotactic sedimentation are weak. We recover isotropy of the active diffusivity tensor and the correct factor of $1/6$ for the dispersal of active swimmers in three dimensions \citep[][]{Berg75}. 

Furthermore, we discover that the linear terms vanish and that up to quadratic order 
\begin{eqnarray}
D_\perp &\sim & \frac{U^2}{D_r}\left[\frac{1}{6} - \frac{\lambda^2}{40} - \frac{\mu^2}{90} + \mathrm{h.o.t.}\right],\\
D_\parallel &\sim & \frac{U^2}{D_r}\left[\frac{1}{6} - \frac{7\lambda^2}{135} -\frac{\lambda\mu}{9} - \frac{2\mu^2}{135} + \mathrm{h.o.t.}\right].
\end{eqnarray}
This explains our previous observation that the diffusivity has a local maximum at $\lambda=0, \ \mu =0$, which can be seen in figure~\ref{fig:diffusivities}. Thus, for small values of $\lambda$ and $\mu$ both mechanisms of chemotactic alignment and indirect chemotactic sedimentation have the effect of suppressing diffusion relative to the mean drift velocity of the swimmers.

\section{Conclusions}

The idea of using autophoresis to design particles capable of independent propulsion and reorientation in the presence of chemical stimuli has gained much attention in recent years, not the least because it eliminates the need for active steering or fine control of external (e.g.~electric or magnetic) fields. In this  paper we  derive a general law for the instantaneous behaviour of a spherical axisymmetric swimmer placed in a uniform chemical gradient, which extends and corrects  results published in \citet{Saha2014}. We also use our framework to calculate the linear and angular velocity of a Janus sphere, which has great relevance for experimental studies. 

The main contribution of the present paper is to obtain a fully analytical solution for a non-trivial transport problem involving a chemically active phoretic swimmer placed in a uniform chemical gradient, a canonical setup which had yet to be  solved in a general form and presented in a pedagogical manner. Furthermore, our systematic analysis of the different sources contributing to chemotaxis could help inform the design of phoretic swimmers in future experiments, as we discussed in the subsection on Janus spheres.

We reinforce the rationale for our first modelling assumption by noting that the regime in which the surface reaction rate is linear in the substrate concentration is the most interesting and relevant case to artificial chemotaxis, because the product problem is sensitive to the substrate gradient in this limit. Our second modelling assumption of small Damk\"{o}hler number could be relaxed, but in this case the substrate problem would no longer have a closed analytical solution and we would have to resort to numerical methods for solving a truncated system. The instantaneous model could be  further extended through the inclusion of advective effects, which have been highlighted by several studies \citep{khair2013, michelin2013c, Lauga14}, or by considering the possible ionic effects recently pointed out by \citet{brown2014}.

In the second half of the paper we present the calculations involved in applying generalised Taylor dispersion theory \citep{Frank91, Frank93} to the artificial chemotaxis of phoretic swimmers, and we obtain very good agreement between the continuum model and numerical  simulations. We observe non-trivial variations of the active diffusivity in the two-dimensional parameter space of rotational P\'{e}clet number and indirect chemotactic index, which creates the possibility of novel pattern formation in systems where these parameters vary with position. The indirect chemotactic index is an added degree of complexity that is new to our continuum model compared to previous studies since it is a direct consequence of the complex chemical dynamics considered in our derivation of the instantaneous behaviour.

The limitations of the present continuum model must however be acknowledged. The main difference between the problem of autophoretic swimmers and that of gyrotactic micro-organisms is that the swimming velocity of the former depends on position, whereas for the latter it is constant. This imposes a further condition on the applicability of the model, namely that the swimming velocity is sufficiently slowly varying in space, which may not be the case in certain practical applications. Furthermore, our model is three-dimensional, whereas many experimental projects are concerned with two-dimensional distributions of autophoretic swimmers above a flat plate.

The model in this paper describes the long-time behaviour of a single phoretic swimmer, and therefore it is valid only for a population of non-interacting autophoretic swimmers. This makes our continuum model unable of explaining the complex collective behaviour observed in dense populations of interacting phoretic swimmers \citep{palacci2013, Saha2014}. Nevertheless, it represents an appropriate starting point for investigating the emergent behaviour of dilute suspensions of chemically active colloidal particles, such as the recent work done on phoretic swimmers ``riding'' active density waves \citep[see][]{geiseler2016}. 

In the phoretic swimming literature, many studies assume   that active swimming leads to an enhanced, but still isotropic, spatial diffusion coefficient, whereas active reorientation leads to a modified rotational diffusion coefficient, and that the two mechanisms are independent of each other. We clearly show in this paper the importance of having an anisotropic diffusivity tensor which encapsulates the effect of both mechanisms, and which represents the correct way to think about the dispersion of autophoretic swimmers. On a phenomenological level, this represents an important contribution of our continuum model to the general understanding of the long-time behaviour of phoretic swimmers placed in a chemical gradient.

\smallskip
We thank the anonymous referees for their useful comments on the early version of our manuscript. This project has received funding from the European Research Council (ERC) under the European Union's Horizon 2020 research and innovation programme  (grant agreement 682754 to EL). This work was also funded by a Summer Research Studentship from Trinity College and a George and Lilian Schiff Studentship (MTC). 

\appendix
\section{}\label{appA}

This appendix contains detailed calculations that complement the subsections \ref{subsec-leading} and \ref{subsec-first} from the derivation of the instantaneous behaviour.

\subsection{Leading order: Linear velocity due to substrate}

To compute the linear velocity due to the leading-order substrate concentration, we must average the slip velocity 
\begin{equation}
\vvb^{s,0}_{slip} = \frac{3\mu_s(\theta)}{2}  \left[(- \nabla s_z^\infty \sin\theta + \nabla s_y^\infty \cos\theta\sin\phi) \skew3\hat{\thb} + \nabla s_y^\infty \cos\phi \ \skew3\hat{\phb} \right] 
\end{equation}
 over the surface of the sphere. The contribution due to the $\skew3\hat{\thb}$ term in (\ref{eq1-solute}) is given by
\begin{equation}
\frac{1}{4\pi} \int_0^{2\pi}\int_0^{\pi} \mu_{s}(\theta) \ \frac{3}{2} (- \nabla s_z^\infty \sin\theta + \nabla s_y^\infty \cos\theta\sin\phi) \vecth \sin\theta \mathrm{d}\theta\mathrm{d}\phi,
\end{equation}
where we have expanded the vector $\skew3\hat{\thb}$ into Cartesian coordinates. The only terms with a non-zero average over azimuthal angle are
\begin{equation}
\frac{3}{8\pi} \int_0^{2\pi}\int_0^{\pi} \sum_{l=0}^{\infty} \mu_{sl}P_l(\cos\theta) (\nabla s_z^\infty \sin^2\theta \skew3\hat{\zzb} + \nabla s_y^\infty \cos^2\theta \sin^2\phi \skew3\hat{\yyb}) \sin\theta \mathrm{d}\theta\mathrm{d}\phi,
\end{equation}
where we have   replaced the mobility $\mu_{s}(\theta)$ by its series representation. Next, we carry out the integral over the azimuthal angle and use a standard change of variable $u=\cos\theta$ for the polar angle to obtain
\begin{equation}
\frac{3}{4} \sum_{l=0}^{\infty} \mu_{sl} \int_{-1}^{+1} P_l(u) \left(\nabla s_z^\infty (1-u^2) \skew3\hat{\zzb} + \frac{1}{2} \nabla s_y^\infty u^2 \skew3\hat{\yyb} \right) \mathrm{d}u.
\end{equation}
Using the classical identities \eqref{eq-Legendre-ortho-cond} and \eqref{eq-A-firstfew} we obtain the final result for the $\skew3\hat{\thb}$ contribution  as
\begin{equation}
(\mu_{s0} - \frac{1}{5} \mu_{s2})\nabla s_z^\infty \skew3\hat{\zzb} + (\frac{1}{4} \mu_{s0} + \frac{1}{10} \mu_{s2}) \nabla s_y^\infty \skew3\hat{\yyb}.
\end{equation}
Following the same principles, the contribution due to the $\skew3\hat{\phb}$ term in (\ref{eq1-solute}) is
\begin{eqnarray}
&~& \frac{1}{4\pi} \int_0^{2\pi} \int_0^{\pi}\ \mu_{s}(\theta) \ \frac{3}{2} \nabla s_y^\infty \cos\phi \vecph \sin\theta \mathrm{d}\theta\mathrm{d}\phi \\
&=& \frac{3}{8\pi} \int_0^{2\pi} \int_0^{\pi} \sum_{l=0}^{\infty} \mu_{sl} P_l(\cos\theta) \nabla s_y^\infty \cos^2\phi \ \skew3\hat{\yyb} \ \sin\theta \mathrm{d}\theta\mathrm{d}\phi \\ 
&=& \frac{3}{8} \sum_{l=0}^{\infty} \mu_{sl} \int_{-1}^{+1} P_l(u)\nabla s_y^\infty \skew3\hat{\yyb} \ \mathrm{d}u \\
&=& \frac{3}{4} \mu_{s0} \nabla s_y^\infty \skew3\hat{\yyb}. 
\end{eqnarray}

As a consequence, the slip velocity averaged over the surface is finally
\begin{equation}
\langle \vvb^{s,0}_{slip}\rangle = (\mu_{s0} - \frac{1}{5} \mu_{s2})\nabla s_z^\infty \skew3\hat{\zzb} + (\mu_{s0} + \frac{1}{10} \mu_{s2}) \nabla s_y^\infty \skew3\hat{\yyb}.
\end{equation}
From   \eqref{eq-linear-velocity} we deduce that the contribution of the substrate to the linear velocity of the particle, at leading order, is 
\begin{equation}
\VVb^{s}_0 = -\left(\mu_{s0}+\frac{1}{10}\mu_{s2}\right) \bnabla s^\infty + \frac{3}{10}\mu_{s2} \ \skew3\hat{\zzb}\skew3\hat{\zzb} \bcdot \bnabla s^\infty.
\end{equation}

\subsection{Leading order: Angular velocity due to substrate}
In order to calculate the angular velocity, we must average the following quantity
\begin{equation}
\hat{\rrb} \wedge \vvb^{s,0}_{slip} =  \frac{3\mu_s(\theta)}{2} \left[(\nabla s_y^\infty \cos\theta\sin\phi - \nabla s_z^\infty \sin\theta) \skew3\hat{\phb} - \nabla s_y^\infty \cos\phi \ \skew3\hat{\thb} \ \right]
\label{eq2-solute}
\end{equation}
over the surface of the sphere. The contribution due to the $\skew3\hat{\phb}$ term in (\ref{eq2-solute}) is
\begin{eqnarray}
&~& \frac{1}{4\pi} \int_0^{2\pi} \int_0^{\pi} \mu_s(\theta) \ \frac{3}{2} (\nabla s_y^\infty \cos\theta\sin\phi - \nabla s_z^\infty \sin\theta) \vecph \sin\theta\mathrm{d}\theta \mathrm{d}\phi \\
&=& \frac{3}{8\pi} \int_0^{2\pi}\int_0^{\pi} \sum_{l=0}^{\infty} \mu_{sl} P_l(\cos\theta) (-\nabla s_y^\infty \cos\theta \sin^2\phi \skew3\hat{\xxb}) \sin\theta\mathrm{d}\theta \mathrm{d}\phi   \\
&=& -\frac{3}{8} \sum_{l=0}^{\infty} \mu_{sl} \int_{-1}^{+1} P_l(u) \nabla s_y^\infty u \ \skew3\hat{\xxb} \ \mathrm{d}u \\
&=& -\frac{1}{4} \mu_{s1} \nabla s_y^\infty \skew3\hat{\xxb} .
\end{eqnarray}
Next, we evaluate the contribution due to the $\skew3\hat{\thb}$ term in (\ref{eq2-solute}) as
\begin{eqnarray}
&~& \frac{1}{4\pi} \int_0^{2\pi} \int_0^{\pi} \mu_s(\theta) \ \frac{3}{2} (-\nabla s_y^\infty \cos\phi) \vecth \sin\theta\mathrm{d}\theta \mathrm{d}\phi \\
&=& \frac{3}{8\pi}  \int_0^{2\pi} \int_0^{\pi} \sum_{l=0}^{\infty} \mu_{sl} P_l(\cos\theta)(-\nabla s_y^\infty \cos\theta\cos^2\phi \skew3\hat{\xxb}) \sin\theta\mathrm{d}\theta\mathrm{d}\phi  \\
&=& -\frac{3}{8} \sum_{l=0}^{\infty} \mu_{sl} \int_{-1}^{+1} P_l(u) \nabla s_y^\infty u \ \skew3\hat{\xxb} \ \mathrm{d}u \\
&=& -\frac{1}{4} \mu_{s1} \nabla s_y^\infty \skew3\hat{\xxb} .
\end{eqnarray}
Adding these two results  gives us
\begin{equation}
\langle \hat{\rrb} \wedge \vvb^{s,0}_{slip} \rangle = - \frac{1}{2} \mu_{s1}\nabla s_y^\infty \skew3\hat{\xxb},
\end{equation}
and from   \eqref{eq-angular-velocity} we deduce that the angular velocity due to the substrate is
\begin{equation}
\omb^{s}_0 =  -\frac{3\mu_{s1}}{4R} \ \skew3\hat{\zzb} \wedge \bnabla s^\infty.
\end{equation}

\subsection{First order: Substrate concentration}

The normal gradient of the substrate concentration from \eqref{eq-s-1-general-form} is
\begin{equation}
\left.\frac{\p s_1}{\p n}\right|_{r=R} = \sum_{l=0}^{\infty} \frac{-(l+1)}{R} \left[A_l P_l(\cos\theta) + B_l P_l^1(\cos\theta) \sin\phi \right].
\end{equation}
We can substitute this result, together with the series representation of $\sigma(\theta)$ and the expression for $s_0$ from \eqref{eq-substrate-concentration}, into the normal-flux boundary condition from \eqref{eq-substrate-first} to get
\begin{multline}
-\frac{D_s}{R}\sum_{l=0}^{\infty} (l+1)\left[A_l P_l(\cos\theta) + B_l P_l^1(\cos\theta) \sin\phi\right] = \\
\kappa_1 \left(s_b (\rrb_c) + \frac{3}{2} \nabla s_z^\infty R \cos\theta + \frac{3}{2} \nabla s_y^\infty R \sin\theta \sin\phi\right) \sum_{j=0}^{\infty} \sigma_j P_j(\cos\theta)
\label{eq-prod-chem-BC}.
\end{multline}
We next separate the components with no $\phi$-dependence by averaging over the azimuthal angle and then take $\int_0^\pi \dots P_k(\cos\theta) \sin\theta\mathrm{d}\theta$ of both sides to get
\begin{multline}
-\frac{D_s}{R}\sum_{l=0}^{\infty} (l+1)A_l \int_{-1}^{+1} P_l(u) P_k(u)\mathrm{d}u  = \\
\kappa_1 \sum_{j=0}^{\infty} \sigma_j \left( s_b (\rrb_c) \int_{-1}^{+1} P_j(u) P_k(u)\mathrm{d}u + \frac{3}{2} \nabla s_z^\infty R \int_{-1}^{+1} u P_j(u) P_k(u)\mathrm{d}u \right).
\end{multline}
Using identities \eqref{eq-Legendre-ortho-cond} and \eqref{eq-uPkPl-integral}, we obtain the $A_k $ set of coefficients as
\begin{equation}
A_k = -\frac{\kappa_1 R}{(k+1)D_s} \left[s_b (\rrb_c) \sigma_k + \frac{3}{2} \nabla s_z^\infty R \left(\frac{k+1}{2k+3} \sigma_{k+1} + \frac{k}{2k-1}\sigma_{k-1}\right)\right].
\label{product-coeff-A}
\end{equation}
We can also separate the $\sin\phi$ components if we multiply equation \eqref{eq-prod-chem-BC} by $\sin\phi$ and average over the azimuthal angle. Then we exploit  the property that $P_l^1(\cos\theta) = -\sin\theta ~ P'_l(\cos\theta)$ and change variables to $u = \cos\theta$ to obtain
\begin{equation}
\frac{D_s}{R}\sum_{l=0}^{\infty} (l+1) B_l P'_l(u) = \frac{3}{2} \kappa_1 \nabla s_y^\infty R \sum_{j=0}^{\infty} \sigma_j P_j(u).
\end{equation}
Taking $\int_{-1}^{+1} \dots (1-u^2) P'_k(u)\mathrm{d}u$ of both sides of this equation, we obtain 
\begin{multline}
\frac{D_s}{R}\sum_{l=0}^{\infty} (l+1) B_l \int_{-1}^{+1} (1-u^2)P'_l(u) P'_k(u)\mathrm{d}u  = \\ \frac{3}{2} \kappa_1\nabla s_y^\infty R \sum_{j=0}^{\infty} \sigma_j \int_{-1}^{+1} (1-u^2) P_j(u)P'_k(u)\mathrm{d}u.
\end{multline}
Finally, using the identities \eqref{eq-1-u2-Pk'-Pl'} and \eqref{eq-1-u2-Pk'-Pl} we obtain the $B_k$ set of coefficients as
\begin{equation}
B_k = -\frac{3\kappa_1 R^2 \nabla s_y^\infty}{2(k+1)D_s} \left(\frac{\sigma_{k+1}}{2k+3} - \frac{\sigma_{k-1}}{2k-1} \right).
\label{product-coeff-B}
\end{equation}

Although the first-order product distribution is now fully determined (since both the $A_k$ and $B_k$ are known),  for the sake of simplicity we will continue the calculations below using the expression
\begin{equation}
s_1 = \sum_{l=0}^{\infty} \left(\frac{R}{r}\right)^{l+1} \left[A_l P_l(\cos\theta) - B_l \sin\theta P'_l(\cos\theta) \sin\phi \right]
\end{equation}
and will substitute the values for $A_l$ and $B_l$ only when necessary.

\subsection{First order: Linear velocity due to substrate}

To calculate the first-order contribution of the substrate problem to the linear velocity, we must average the slip velocity
\begin{multline}
\vvb^{s,1}_{\text{slip}} = \frac{\mu_s(\theta)}{R} \sum_{l=0}^{\infty} \left(- A_l\sin\theta P'_l(\cos\theta) -B_l \cos\theta P'_l(\cos\theta) \sin\phi\right. \\ + \left.B_l \sin^2\theta P''_l(\cos\theta)\sin\phi\right)\skew3\hat{\thb} + \frac{\mu_s(\theta)}{R} \sum_{l=0}^{\infty} \left(-B_l P'_l(\cos\theta)\cos\phi\right) \skew3\hat{\phb}. 
\label{product-eq1}
\end{multline}
over the surface of the swimmer.

The contribution to $\langle \vvb^{s,1}_{\text{slip}} \rangle$ due to the $\skew3\hat{\thb}$ terms in (\ref{product-eq1}) is  given by
\begin{multline}
\frac{1}{4\pi} \int_0^{2\pi}\int_0^{\pi}\mu_{s}(\theta) \frac{1}{R}\sum_{l=0}^{\infty} \left[- A_l\sin\theta P'_l(\cos\theta) -B_l \cos\theta P'_l(\cos\theta)\sin\phi\right. \\ \left. + B_l \sin^2\theta P''_l(\cos\theta)\sin\phi \right] \vecth \sin\theta~\mathrm{d}\theta\mathrm{d}\phi .
\end{multline}
We can first  average over the azimuthal angle and perform the usual change of variable for the polar angle in order to obtain the average
\begin{multline}\label{2.55}
\frac{1}{2R} \sum_{m,l=0}^\infty\mu_{sm}A_l \int_{-1}^{+1} (1-u^2) P_m(u) P'_l(u) \mathrm{d}u \ \skew3\hat{\zzb} \\
+ \frac{1}{4R} \sum_{m,l=0}^\infty\mu_{sm}B_l \left[ \int_{-1}^{+1} \ u(1-u^2) P_m(u) P''_l(u)\mathrm{d}u - \int_{-1}^{+1} u^2 P_m(u) P'_l(u) \mathrm{d}u \right] \skew3\hat{\yyb},
\end{multline}
where we have replaced $\mu_{s}(\theta)$ by its series representation.

The $\skew3\hat{\zzb}$ component of the integral in \eqref{2.55} is straightforward, but for the $\skew3\hat{\yyb}$ component we need to do an integration by parts on the first integral to bring it to the form 
\begin{equation}
-\int_{-1}^{+1} \left[u(1-u^2) \ P'_m(u) + (1-3u^2)P_m(u) \right] P'_l(u)\mathrm{d}u.
\end{equation} 
As a result, the contribution to linear velocity due to the $\skew3\hat{\thb}$ terms in (\ref{product-eq1}) is given by
\begin{multline}
\quad \frac{1}{2R} \sum_{m,l=0}^\infty\mu_{sm}A_l \int_{-1}^{+1} (1-u^2) P_m(u) P'_l(u)\mathrm{d}u \  \skew3\hat{\zzb} \\
+ \frac{1}{4R} \sum_{m,l=0}^\infty\mu_{sm}B_l \left[ -\int_{-1}^{+1} u(1-u^2) P'_m(u) P'_l(u) \mathrm{d}u -2\int_{-1}^{+1} (1-u^2) P_m(u) P'_l(u)\mathrm{d}u  \right. \\ \left. + \int_{-1}^{+1} P_m(u) P'_l(u) \mathrm{d}u \right] \skew3\hat{\yyb}.
\end{multline}
We can evaluate explicitly  all but the last of those integrals, but fortunately the contribution to $\langle \vvb^{s,1}_{\text{slip}} \rangle$ due to the $\skew3\hat{\phb}$ term in (\ref{product-eq1}) is exactly
\begin{multline}
\frac{1}{4\pi} \int_0^{2\pi} \int_0^{\pi} \mu_s(\theta) \ \frac{1}{R} \sum_{l=0}^{\infty} \left(-B_l P'_l(\cos\theta)\cos\phi\right)\vecph \sin\theta\mathrm{d}\theta\mathrm{d}\phi \\
= -\frac{1}{4R} \sum_{m,l=0}^\infty\mu_{sm}B_l \int_{-1}^{+1} P_m(u) P'_l(u) \mathrm{d}u \ \skew3\hat{\yyb}, 
\end{multline}
so the undetermined integrals exactly cancel out. Using identities \eqref{eq-1-u2-Pk'-Pl}, \eqref{eq-(1-u2)uPk'Pl'} and equation  
\eqref{eq-linear-velocity},   we deduce that the first-order substrate contribution to linear velocity is given by
\begin{multline}
\VVb^s_1 = \frac{1}{R} \sum_{l=0}^{\infty} \frac{l(l+1)}{2l+1} A_l \left( \frac{\mu_{s,l+1}}{2l+3}-\frac{\mu_{s,l-1}}{2l-1}\right) \skew3\hat{\zzb} \\ + \frac{1}{2R}  \sum_{l=0}^{\infty} \frac{l(l+1)}{2l+1} B_l \left( \frac{l\mu_{s,l+1}}{2l+3}+\frac{(l+1)\mu_{s,l-1}}{2l-1}\right) \skew3\hat{\yyb} .
\end{multline}
If we substitute the values of the coefficients $A_l$ and $B_l$ from   (\ref{product-coeff-A}) and (\ref{product-coeff-B}), we get
\begin{multline}
\VVb^s_1 = -\frac{\kappa_1 s_b (\rrb_c)}{D_s} \sum_{l=0}^{\infty} \left(\frac{l}{2l+1}\right) \   \sigma_l \left( \frac{\mu_{s,l+1}}{2l+3}-\frac{\mu_{s,l-1}}{2l-1}\right) \skew3\hat{\zzb} \\
- \frac{3\kappa_1 R}{2D_s}\sum_{l=0}^{\infty} \left(\frac{l}{2l+1}\right) \left(\frac{l+1}{2l+3} \sigma_{l+1} + \frac{l}{2l-1}\sigma_{l-1}\right) \left( \frac{\mu_{s,l+1}}{2l+3}-\frac{\mu_{s,l-1}}{2l-1}\right) \nabla s_z^\infty  \skew3\hat{\zzb} \\
- \frac{3\kappa_1 R}{4D_s} \sum_{l=0}^{\infty} \left(\frac{l}{2l+1}\right) \left(\frac{\sigma_{l+1}}{2l+3} - \frac{\sigma_{l-1}}{2l-1}\right) \left(\frac{l+1}{2l-1}\mu_{s,l-1}+\frac{l}{2l+3}\mu_{s,l+1}\right) \nabla s_y^\infty \skew3\hat{\yyb}.
\label{eq-needed-for-janus} 
\end{multline}
We can further manipulate the last two terms in order to get an expression in terms of $\bnabla s^\infty$ and $\skew3\hat{\zzb}\skew3\hat{\zzb} \bcdot \bnabla s^\infty$ for the linear velocity as
\begin{multline}
\VVb^{s}_1 = -\frac{\kappa_1 s_b (\rrb_c)}{D_s} \sum_{l=1}^{\infty} \left(\frac{l}{2l+1}\right)\sigma_l \left(\frac{\mu_{s,l+1}}{2l+3}-\frac{\mu_{s,l-1}}{2l-1}\right) \skew3\hat{\zzb} \\
- \frac{3 \kappa_1 R}{4 D_s}\sum_{l=1}^{\infty} \left(\frac{l}{2l+1}\right)\left(\frac{\sigma_{l+1}}{2l+3} - \frac{\sigma_{l-1}}{2l-1} \right)\left(\frac{l+1}{2l-1}\mu_{s,l-1}+\frac{l}{2l+3}\mu_{s,l+1}\right) \bnabla s^\infty \\
+ \frac{3 \kappa_1 R}{4 D_s}\sum_{l=1}^{\infty} \left(\frac{l}{2l+1}\right)\left(\frac{3(l+1)\sigma_{l+1}\mu_{s,l-1}}{(2l+3)(2l-1)}-\frac{(l+2)\sigma_{l+1}\mu_{s,l+1}}{(2l+3)^2}\right. \\ +\left. \frac{(l-1)\sigma_{l-1}\mu_{s,l-1}}{(2l-1)^2} - \frac{3l\sigma_{l-1}\mu_{s,l+1}}{(2l-1)(2l+3)}\right) \skew3\hat{\zzb}\skew3\hat{\zzb}\bcdot\bnabla s^\infty.
\end{multline}

\subsection{First order: Angular velocity due to substrate}
We next need to evaluate $\langle \hat{\rrb} \wedge \vvb^{s,1}_{\text{slip}} \rangle$ for the angular velocity. This angular integral  
is equal to
\begin{multline}
\frac{1}{4\pi} \int_0^{2\pi} \int_0^{\pi} \mu_s(\theta) \left[ \frac{1}{R} \sum_{l=0}^{\infty} \left(- A_l\sin\theta P'_l(\cos\theta) -B_l \cos\theta P'_l(\cos\theta) \sin\phi\right.\right. \\ + \left.\left.B_l \sin^2\theta P''_l(\cos\theta)\sin\phi\right)\skew3\hat{\phb} + \frac{1}{R} \sum_{l=0}^{\infty} B_l P'_l(\cos\theta)\cos\phi \ \skew3\hat{\thb}\right] \sin\theta\mathrm{d}\theta\mathrm{d}\phi. 
\label{eq-ang-vel-prod-integral}
\end{multline}
The $\skew3\hat{\phb}$ contribution of this integral is
\begin{equation}
\frac{1}{4R} \sum_{m,l=0}^\infty \mu_{sm} B_l \left[ \int_{-1}^{+1} uP_m(u)P'_l(u)\mathrm{d}u - \int_{-1}^{+1} (1-u^2)P_m(u) P''_l(u)\mathrm{d}u \right] \skew3\hat{\xxb} . 
\end{equation}
We may integrate by parts the second integral in order to get
\begin{equation}
\frac{1}{4R} \sum_{m,l=0}^\infty \mu_{sm} B_l \left[ -\int_{-1}^{+1} uP_m(u)P'_l(u)\mathrm{d}u  + \int_{-1}^{+1} (1-u^2)P'_m(u) P'_l(u)\mathrm{d}u\right] \skew3\hat{\xxb}.
\end{equation}
The $\skew3\hat{\thb}$ contribution to the integral in \eqref{eq-ang-vel-prod-integral}
is simply given by
\begin{equation}
\frac{1}{4R} \sum_{m,l=0}^\infty \mu_{sm} B_l \int_{-1}^{+1} uP_m(u)P'_l(u) \mathrm{d}u \  \skew3\hat{\xxb},
\end{equation}
so by adding the two angular contributions and using identity \eqref{eq-1-u2-Pk'-Pl'} we obtain the surface average
\begin{equation}
\langle \hat{\rrb} \wedge \vvb^{s,1}_{\text{slip}} \rangle = \frac{1}{2R} \sum_{l=0}^{\infty} \frac{l(l+1)}{2l+1} \mu_{sl} B_l \ \skew3\hat{\xxb}.
\end{equation}
Given \eqref{eq-angular-velocity}, the angular velocity due to the first-order substrate concentration takes the final form 
\begin{equation}
\omb^{s}_1= -\frac{9\kappa_1}{8 D_s} \sum_{l=1}^{\infty} \left(\frac{l}{2l+1}\right)\mu_{sl} \left(\frac{\sigma_{l+1}}{2l+3}-\frac{\sigma_{l-1}}{2l-1}\right) ~ \skew3\hat{\zzb} \wedge \bnabla s^\infty 
\end{equation}
where we have substituted the value of $B_l$ from   \eqref{product-coeff-B}.

\section{}\label{appB}

This appendix contains a list of all the properties and integral identities  of Legendre polynomials that we have used in the rest of the paper. These are:
\begin{itemize}
\item{
The differential equation satisfied by Legendre polynomials,
\begin{equation}
\left((1-u^2)P_k'(u)\right)'+k(k+1)P_k(u) = 0;
\label{eq-first-basic-prop}
\end{equation}
}
\item{
The orthogonality condition for Legendre polynomials,
\begin{equation}
\int_{-1}^{+1} P_k(u) P_l(u) \mathrm{d}u = \frac{2}{2k+1} \delta_{lk};
\label{eq-Legendre-ortho-cond}
\end{equation}
}
\item{
Two useful recurrence relations involving Legendre polynomials and their derivatives,
\begin{eqnarray}
(2k+1)uP_k &=& (k+1)P_{k+1} +  kP_{k-1},\\
(1-u^2)P'_k &=& k \left(P_{k-1}-uP_k\right);
\label{eq-lat-basic-prop}
\end{eqnarray} 
}
\item{
The first three Legendre polynomials,
\begin{equation}
P_0(u) = 1, \quad P_1(u) = u, \quad P_2(u) = \frac{1}{2}(3u^2-1);
\label{eq-A-firstfew}
\end{equation}
} 
\item{
A series of integral identities which can be derived from the basic properties \eqref{eq-first-basic-prop}-\eqref{eq-lat-basic-prop}:
\begin{eqnarray}
\int_{-1}^{+1} u P_k(u) P_l(u) \mathrm{d}u &=& \frac{2}{2k+1} \left(\frac{k+1}{2k+3}\delta_{l,k+1}+\frac{k}{2k-1}\delta_{l,k-1}\right), \label{eq-uPkPl-integral}\\
\int_{-1}^{+1} (1-u^2) P'_k(u) P'_l(u) \mathrm{d}u &=& \frac{2k(k+1)}{2k+1} \delta_{lk}, \label{eq-1-u2-Pk'-Pl'} \\ 
\int_{-1}^{+1} (1-u^2) P'_k(u) P_l(u) \mathrm{d}u &=& \frac{2k(k+1)}{2k+1} \left( \frac{\delta_{l,k-1}}{2k-1} - \frac{\delta_{l,k+1}}{2k+3}\right), \label{eq-1-u2-Pk'-Pl}\\
\int_{-1}^{+1} (1-u^2)u P'_k(u) P'_l(u) \mathrm{d}u &=& \frac{2k(k+1)}{2k+1} \left( \frac{k+2}{2k+3} \delta_{l,k+1}+\frac{k-1}{2k-1} \delta_{l,k-1}\right), \label{eq-(1-u2)uPk'Pl'} \\
\int_{-1}^{+1} (1-u^2)u P_k(u) P'_l(u) \mathrm{d}u &=& \frac{2(k+1)(k+2)(k+3)}{(2k+1)(2k+3)(2k+5)} \delta_{l,k+2}\nonumber\\
&+& \frac{2k(k+1)}{(2k-1)(2k+1)(2k+3)} \delta_{lk}\nonumber\\
&-& \frac{2(k-2)(k-1)k}{(2k-3)(2k-1)(2k+1)} \delta_{l,k-2}, \label{eq-(1-u2)uPk_Pl'}\\
\int_{-1}^{+1} (1-u^2)^2 P_k(u) P''_l(u) \mathrm{d}u &=& \frac{2(k+1)(k+2)(k+3)(k+4)}{(2k+1)(2k+3)(2k+5)} \delta_{l,k+2}\nonumber\\
&-& \frac{4(k-1)k(k+1)(k+2)}{(2k-1)(2k+1)(2k+3)} \delta_{lk}\nonumber\\
&+& \frac{2(k-3)(k-2)(k-1)k}{(2k-3)(2k-1)(2k+1)} \delta_{l,k-2}\label{eq-(1-u2)2_Pk_Pl''};
\end{eqnarray}
}
\item{
And a useful half-integral for our calculations on the Janus sphere:
\begin{equation}
  \int_0^1 P_k(u) \mathrm{d}u = \left\{
    \begin{array}{ll}
       1, & k=0 \\[2pt]
       0, & k=2,4,6,\dots \\[2pt]
       (-1)^{(m-1)/2}\frac{m!!}{m (m+1)!!},& k=1,3,5,\dots
    \end{array} \right.
    \label{eq-half-integral}
\end{equation}
}
\end{itemize}

\section{}\label{appC}

Expressions obtained from the GTD model for $D_\perp$ and $D_\parallel$ by inverting the truncated 2-by-2 systems corresponding to equations \eqref{eq:pentadiagonal} and \eqref{eq:tridiagonal}:

\begin{multline}
D_\perp = \frac{U^2 D_r^{-1}}{3\lambda^4(20+\lambda^2)}\left[-5\lambda^2(9+\lambda^2) -24\lambda(5+2\lambda^2)\mu + (120+41\lambda^2-3\lambda^4)\mu^2 \right. \\ \left. + \left(45\lambda^3+8\lambda^2(15+\lambda^2)\mu - \lambda (120+\lambda^2)\mu^2\right)\coth\lambda \right]
\end{multline}

\begin{multline}
D_\parallel = \frac{U^2 D_r^{-1}}{6\lambda^4(15+\lambda^2)}\left[15\lambda^2(5+4\lambda^2) + 2\lambda(255\lambda+133\lambda^3+8\lambda^5)\mu -4 (105+43\lambda^2-\lambda^4)\mu^2 \right. \\ \left. +\left(-15\lambda^3+20\lambda^2(-33+5\lambda-7\lambda^2)\mu + 12\lambda(40+5\lambda^2-\lambda^4)\mu^2 \right)\coth\lambda \right. \\ \left. +6\lambda^2(\lambda-2\mu)(5\mu-10\lambda-\lambda^2\mu)\coth^2\lambda \right]
\end{multline}

\bibliographystyle{jfm}
\bibliography{artif_chemo_refs}

\begin{thebibliography}{51}
\expandafter\ifx\csname natexlab\endcsname\relax\def\natexlab#1{#1}\fi

\bibitem[Agudo-Canalejo {\em et~al.\/}(2018)Agudo-Canalejo, Illien \&
  Golestanian]{Agudo2018}
{\sc Agudo-Canalejo, J., Illien, P. \& Golestanian, R.} 2018 Phoresis and
  enhanced diffusion compete in enzyme chemotaxis. {\em Nano Lett.\/} {\bf 18},
  2711--2717.

\bibitem[Anderson(1989)]{Anderson89}
{\sc Anderson, J.L.} 1989 Colloid transport by interfacial forces. {\em Annu.
  Rev. Fluid Mech.\/} {\bf 21}, 61--99.

\bibitem[Batchelor(1976)]{Batchelor76}
{\sc Batchelor, G.K.} 1976 Brownian diffusion of particles with hydrodynamic
  interaction. {\em J. Fluid Mech.\/} {\bf 74}, 1--29.

\bibitem[Bearon {\em et~al.\/}(2012)Bearon, Bees \& Croze]{Bees12}
{\sc Bearon, R.N., Bees, M.A. \& Croze, O.A.} 2012 Biased swimming cells do not
  disperse in pipes as tracers: A population model based on microscale
  behaviour. {\em Phys. Fluids\/} {\bf 24}, 121902.

\bibitem[Berg(1975)]{Berg75}
{\sc Berg, H.C.} 1975 Chemotaxis in bacteria. {\em Annu. Rev. Biophys.
  Bioeng.\/} {\bf 4}, 119--136.

\bibitem[Bickel {\em et~al.\/}(2013)Bickel, Majee \& W\"urger]{bickel2013}
{\sc Bickel, T., Majee, A. \& W\"urger, A.} 2013 Flow pattern in the vicinity
  of self-propelling hot {J}anus particles. {\em Phys. Rev. E\/} {\bf 88},
  012301.

\bibitem[Bickel {\em et~al.\/}(2014)Bickel, Zecua \& W\"urger]{bickel2014}
{\sc Bickel, T., Zecua, G. \& W\"urger, A.} 2014 Polarization of active {J}anus
  particles. {\em Phys. Rev. E\/} {\bf 89}, 050303(R).

\bibitem[Brady(2011)]{brady2011}
{\sc Brady, J.} 2011 Particle motion driven by solute gradients with
  application to autonomous motion: continuum and colloidal perspectives. {\em
  J. Fluid Mech.\/} {\bf 667}, 216--259.

\bibitem[Brown \& Poon(2014)]{brown2014}
{\sc Brown, A. \& Poon, W.} 2014 Ionic effects in self-propelled {P}t-coated
  {J}anus swimmers. {\em Soft Matter\/} {\bf 10}, 4016--4027.

\bibitem[C{\'o}rdova-Figueroa \& Brady(2008)]{cordova2008}
{\sc C{\'o}rdova-Figueroa, U.~M. \& Brady, J.~F.} 2008 Osmotic propulsion: The
  osmotic motor. {\em Phys. Rev. Lett.\/} {\bf 100}, 158303.

\bibitem[C{\'o}rdova-Figueroa {\em et~al.\/}(2013)C{\'o}rdova-Figueroa, Brady
  \& Shklyaev]{cordova2013}
{\sc C{\'o}rdova-Figueroa, U.~M., Brady, J.~F. \& Shklyaev, S.} 2013 Osmotic
  propulsion of colloidal particles via constant surface flux. {\em Soft
  Matter\/} {\bf 9}, 6382--6390.

\bibitem[Ebbens {\em et~al.\/}(2014)Ebbens, Gregory, Dunderdale, Howse,
  Ibrahim, Liverpool \& Golestanian]{ebbens2014}
{\sc Ebbens, S., Gregory, D.~A., Dunderdale, G., Howse, J.~R., Ibrahim, Y.,
  Liverpool, T.~B. \& Golestanian, R.} 2014 Electrokinetic effects in catalytic
  platinum-insulator {Janus} swimmers. {\em Eur. Phys. Lett.\/} {\bf 106},
  58003.

\bibitem[Ebbens {\em et~al.\/}(2012)Ebbens, Tu, Howse \&
  Golestanian]{ebbens2012}
{\sc Ebbens, S., Tu, M.-H., Howse, J.~R. \& Golestanian, R.} 2012 Size
  dependence of the propulsion velocity for catalytic {Janus}-sphere swimmers.
  {\em Phys. Rev. E\/} {\bf 85}, 020401.

\bibitem[Ebbens \& Howse(2011)]{Ebbens2011}
{\sc Ebbens, S.~J. \& Howse, J.~R.} 2011 Direct observation of the direction of
  motion for spherical catalytic swimmers. {\em Langmuir\/} {\bf 27},
  12293--12296.

\bibitem[Einstein(1905)]{Einstein1905}
{\sc Einstein, A.} 1905 \"{U}ber die von der molekularkinetischen {T}heorie der
  {W}\"{a}rme geforderte {B}ewegung von in ruhenden {F}l\"{u}ssigkeiten
  suspendierten {T}eilchen. {\em Ann. Phys.\/} {\bf 322}, 549--560.

\bibitem[Frankel \& Brenner(1991)]{Frank91}
{\sc Frankel, I. \& Brenner, H.} 1991 Generalized {Taylor} dispersion phenomena
  in unbounded homogeneous shear flows. {\em J. Fluid Mech.\/} {\bf 230},
  147--181.

\bibitem[Frankel \& Brenner(1993)]{Frank93}
{\sc Frankel, I. \& Brenner, H.} 1993 {Taylor} dispersion of orientable
  {Brownian} particles in unbounded homogeneous shear flows. {\em J. Fluid
  Mech.\/} {\bf 255}, 129--156.

\bibitem[Geiseler {\em et~al.\/}(2016)Geiseler, Hanggi, Marchesoni, Mulhern \&
  Savel'ev]{geiseler2016}
{\sc Geiseler, A., Hanggi, P., Marchesoni, F., Mulhern, C. \& Savel'ev, S.}
  2016 Chemotaxis of artificial microswimmers in active density waves. {\em
  Phys. Rev. E\/} {\bf 94}, 012613.

\bibitem[Golestanian(2012)]{golestanian2012}
{\sc Golestanian, R.} 2012 Collective behavior of thermally active colloids.
  {\em Phys. Rev. Lett.\/} {\bf 108}, 3.

\bibitem[Golestanian {\em et~al.\/}(2005)Golestanian, Liverpool \&
  Ajdari]{golestanian2005}
{\sc Golestanian, R., Liverpool, T.~B. \& Ajdari, A.} 2005 Propulsion of a
  molecular machine by asymmetric distribution of reaction products. {\em Phys.
  Rev. Lett.\/} {\bf 94}, 220801.

\bibitem[Golestanian {\em et~al.\/}(2007)Golestanian, Liverpool \&
  Ajdari]{Golestanian07}
{\sc Golestanian, R., Liverpool, T.~B. \& Ajdari, A.} 2007 Designing phoretic
  micro- and nanoswimmers. {\em New J. Phys.\/} {\bf 9}, 126.

\bibitem[Hill \& Bees(2002)]{Hill02}
{\sc Hill, N.A. \& Bees, M.A.} 2002 {Taylor} dispersion of gyrotactic swimming
  micro-organisms in a linear flow. {\em Phys. Fluids.\/} {\bf 14}, 2598--2605.

\bibitem[Howse {\em et~al.\/}(2007)Howse, Jones, Ryan, Gough, Vafabakhsh \&
  Golestanian]{Howse2007}
{\sc Howse, J.~R., Jones, R. A.~L., Ryan, A.~J., Gough, T., Vafabakhsh, R. \&
  Golestanian, R.} 2007 Self-motile colloidal particles: From directed
  propulsion to random walk. {\em Phys. Rev. Lett.\/} {\bf 99}, 048102.

\bibitem[Izri {\em et~al.\/}(2014)Izri, van~der Linden, Michelin \&
  Dauchot]{izri2014}
{\sc Izri, Z., van~der Linden, M.~N., Michelin, S. \& Dauchot, O.} 2014
  Self-propulsion of pure water droplets by spontaneous {M}arangoni stress
  driven motion. {\em Phys. Rev. Lett.\/} {\bf 113}, 248302.

\bibitem[Jiang {\em et~al.\/}(2010)Jiang, Yoshinaga \& Sano]{jiang2010}
{\sc Jiang, H.-R., Yoshinaga, N. \& Sano, M.} 2010 Active motion of a {J}anus
  particle by self-thermophoresis in a defocused laser beam. {\em Phys. Rev.
  Lett.\/} {\bf 105}, 268302.

\bibitem[J{\"u}licher \& Prost(2009)]{julicher2009}
{\sc J{\"u}licher, F. \& Prost, J.} 2009 Generic theory of colloidal transport.
  {\em Eur. Phys. J. E\/} {\bf 29}, 27--36.

\bibitem[Khair(2013)]{khair2013}
{\sc Khair, A.~S.} 2013 Diffusiophoresis of colloidal particles in neutral
  solute gradients at finite {P}\'eclet number. {\em J. Fluid Mech.\/} {\bf
  731}, 64--94.

\bibitem[Lauga(2016)]{Lauga2016}
{\sc Lauga, E.} 2016 Bacterial hydrodynamics. {\em Annu. Rev. Fluid Mech.\/}
  {\bf 48}, 105--130.

\bibitem[Lauga \& Powers(2009)]{Lauga2009}
{\sc Lauga, E. \& Powers, T.R.} 2009 The hydrodynamics of swimming
  micro-organisms. {\em Rep. Prog. Phys.\/} {\bf 72}, 096601.

\bibitem[Manela \& Frankel(2003)]{Frank03}
{\sc Manela, A. \& Frankel, I.} 2003 Generalized {Taylor} dispersion in
  suspensions of gyrotactic swimming micro-organisms. {\em J. Fluid Mech.\/}
  {\bf 490}, 99--127.

\bibitem[Michelin \& Lauga(2014)]{Lauga14}
{\sc Michelin, S. \& Lauga, E.} 2014 Phoretic self-propulsion at finite
  {P\'{e}clet} numbers. {\em J. Fluid Mech.\/} {\bf 747}, 572–--604.

\bibitem[Michelin \& Lauga(2015)]{Lauga15}
{\sc Michelin, S. \& Lauga, E.} 2015 Autophoretic locomotion from geometric
  asymmetry. {\em Eur. Phys. J. E\/} {\bf 38}, 7.

\bibitem[Michelin {\em et~al.\/}(2013)Michelin, Lauga \&
  Bartolo]{michelin2013c}
{\sc Michelin, S., Lauga, E. \& Bartolo, D.} 2013 Spontaneous autophoretic
  motion of isotropic particles. {\em Phys. Fluids\/} {\bf 25}, 061701.

\bibitem[Nelson {\em et~al.\/}(2010)Nelson, Kaliakatsos \& Abbott]{Nelson2010}
{\sc Nelson, B.~J., Kaliakatsos, I.~K. \& Abbott, J.~J.} 2010 Microrobots for
  minimally invasive medicine. {\em Annu. Rev. Biomed. Eng.\/} {\bf 12},
  55--85.

\bibitem[Nelson(2008)]{nelson2008}
{\sc Nelson, P.C.} 2008 {\em Biological {P}hysics: {E}nergy, {I}nformation,
  {L}ife\/}. New York, N.Y.: W.H. Freeman.

\bibitem[Palacci {\em et~al.\/}(2014)Palacci, Sacanna, Kim, Yi, Pine \&
  Chaikin]{palacci2014}
{\sc Palacci, J., Sacanna, S., Kim, S.-H., Yi, G.-R., Pine, D.~J. \& Chaikin,
  P.~M.} 2014 Light-activated self-propelled colloids. {\em Philos. Trans.
  Royal Soc. A\/} {\bf 372}, 2029.

\bibitem[Palacci {\em et~al.\/}(2013)Palacci, Sacanna, Steinberg, Pine \&
  Chaikin]{palacci2013}
{\sc Palacci, J., Sacanna, S., Steinberg, A.~P., Pine, D.~J. \& Chaikin, P.~M.}
  2013 Living crystals of light-activated colloidal surfers. {\em Science\/}
  {\bf 339}, 936--940.

\bibitem[Paxton {\em et~al.\/}(2004)Paxton, Kistler, Olmeda, Sen, St.~Angelo,
  Cao, Mallouk, Lammert \& Crespi]{paxton2004}
{\sc Paxton, W.~F., Kistler, K.~C., Olmeda, C.~C., Sen, A., St.~Angelo, S.~K.,
  Cao, Y., Mallouk, T.~E., Lammert, P.~E. \& Crespi, V.~H.} 2004 Catalytic
  nanomotors: Autonomous movement of striped nanorods. {\em J. Am. Chem.
  Soc.\/} {\bf 126}, 13424--13431.

\bibitem[Pedley \& Kessler(1990)]{Pedley90}
{\sc Pedley, T.J. \& Kessler, J.O.} 1990 A new continuum model for suspensions
  of gyrotactic micro-organisms. {\em J. Fluid Mech.\/} {\bf 212}, 155--182.

\bibitem[Pohl \& Stark(2014)]{pohl2014}
{\sc Pohl, O. \& Stark, H.} 2014 Dynamic clustering and chemotactic collapse of
  self-phoretic active particles. {\em Phys. Rev. Lett.\/} {\bf 112}, 238303.

\bibitem[Popescu {\em et~al.\/}(2010)Popescu, Dietrich, Tasinkevych \&
  Ralston]{popescu2010}
{\sc Popescu, M.~N., Dietrich, S., Tasinkevych, M. \& Ralston, J.} 2010
  Phoretic motion of spheroidal particles due to self-generated solute
  gradients. {\em Eur. Phys J. E\/} {\bf 31}, 351--367.

\bibitem[Popescu {\em et~al.\/}(2011)Popescu, Tasinkevych \&
  Dietrich]{popescu2011}
{\sc Popescu, M.~N., Tasinkevych, M. \& Dietrich, S.} 2011 Pulling and pushing
  a cargo with a catalytically active carrier. {\em Eur. Phys. Lett.\/} {\bf
  95}, 28004.

\bibitem[Sabass \& Seifert(2012)]{sabass2012}
{\sc Sabass, B. \& Seifert, U.} 2012 Dynamics and efficiency of a
  self-propelled, diffusiophoretic swimmer. {\em J. Chem. Phys.\/} {\bf 136},
  064508.

\bibitem[Saha {\em et~al.\/}(2014)Saha, Golestanian \& Ramaswamy]{Saha2014}
{\sc Saha, S., Golestanian, R. \& Ramaswamy, S.} 2014 Clusters, asters, and
  collective oscillations in chemotactic colloids. {\em Phys. Rev. E\/} {\bf
  89}, 062316.

\bibitem[Saragosti {\em et~al.\/}(2012)Saragosti, Silberzan \&
  Buguin]{saragosti12}
{\sc Saragosti, J., Silberzan, P. \& Buguin, A.} 2012 Modeling \textit{E. coli}
  tumbles by rotational diffusion. {I}mplications for chemotaxis. {\em PLoS
  One\/} {\bf 7}, 1--6.

\bibitem[Schmitt \& Stark(2013)]{schmitt2013}
{\sc Schmitt, M. \& Stark, H.} 2013 Swimming active droplet: A theoretical
  analysis. {\em Eur. Phys. Lett.\/} {\bf 101}, 44008.

\bibitem[Shklyaev {\em et~al.\/}(2014)Shklyaev, Brady \&
  C{\'o}rdova-Figueroa]{shklyaev2014}
{\sc Shklyaev, S., Brady, J.~F. \& C{\'o}rdova-Figueroa, U.~M.} 2014
  Non-spherical osmotic motor: Chemical sailing. {\em J. Fluid Mech.\/} {\bf
  748}, 488--520.

\bibitem[von Smoluchowski(1906)]{Smolu1906}
{\sc von Smoluchowski, M.} 1906 Zur kinetischen {T}heorie der {B}rownschen
  {M}olekularbewegung und der {S}uspensionen. {\em Ann. Phys.\/} {\bf 326},
  756--780.

\bibitem[Stone \& Samuel(1996)]{Stone96}
{\sc Stone, H.~A. \& Samuel, A. D.~T.} 1996 Propulsion of microorganisms by
  surface distortions. {\em Phys. Rev. Lett.\/} {\bf 77}, 4102--4104.

\bibitem[Thutupalli {\em et~al.\/}(2011)Thutupalli, Seemann \&
  Herminghaus]{thutupalli2011}
{\sc Thutupalli, S., Seemann, R. \& Herminghaus, S.} 2011 Swarming behavior of
  simple model squirmers. {\em New J. Phys.\/} {\bf 13}, 073021.

\bibitem[Wang {\em et~al.\/}(2013)Wang, Duan, Sen \& Mallouk]{Wang2013}
{\sc Wang, W., Duan, W., Sen, A. \& Mallouk, T.~E.} 2013 Catalytically powered
  dynamic assembly of rod-shaped nanomotors and passive tracer particles. {\em
  Proc. Natl. Acad. Sci.\/} {\bf 110}, 17744--17749.

\end{thebibliography}

\end{document}